\documentclass[prd,superscriptaddress,twocolumn,showpacs,dvips]{revtex4}
\usepackage{feynmp}
\usepackage{amssymb}
\usepackage{epsfig}
\usepackage{graphicx}
\usepackage{subfigure}
\usepackage{hyperref}
\usepackage{bbm}
\usepackage{color}

\begin{document}

\title{Infrared behavior of dynamical fermion mass generation in QED$_{3}$}

\author{Jing-Rong Wang}
\affiliation{High Magnetic Field Laboratory, Hefei Institutes of
Physical Science, Chinese Academy of Sciences, Hefei 230031, P. R.
China}

\author{Guo-Zhu Liu}
\altaffiliation{Corresponding author: gzliu@ustc.edu.cn}
\affiliation{Department of Modern Physics, University of Science and
Technology of China, Hefei, Anhui 230026, P. R. China}

\author{Chang-Jin Zhang}
\altaffiliation{Corresponding author: zhangcj@hmfl.ac.cn}
\affiliation{High Magnetic Field Laboratory, Hefei Institutes of
Physical Science, Chinese Academy of Sciences, Hefei 230031, P. R.
China}

\begin{abstract}
Extensive investigations show that QED$_{3}$ exhibits dynamical
fermion mass generation at zero temperature when the fermion flavor
$N$ is sufficiently small. However, it seems difficult to extend the
theoretical analysis to finite temperature. We study this problem by
means of Dyson-Schwinger equation approach after considering the
effect of finite temperature or disorder-induced fermion damping.
Under the widely used instantaneous approximation, the dynamical
mass displays an infrared divergence in both cases. We then adopt a
new approximation that includes an energy-dependent gauge boson
propagator and obtain results for dynamical fermion mass that do not
contain infrared divergence. The validity of the new approximation
is examined by comparing to the well-established results obtained at
zero temperature.
\end{abstract}


\maketitle

%%%%%%%%%%%%%%%%%%%%%%%%%%%%%Main Body%%%%%%%%%%%%%%%%%%%%%%%%%%%%%%%%%%%%%

\section{Introduction}

It is well established that (3+1)-dimensional quantum
electrodynamics (QED) can describe the electromagnetic interaction
between charged elementary particles with very high precision. QED
defined on (2+1)-dimensional space-time, dubbed QED$_{3}$, is safer
in the high-energy region than its (3+1)-dimensional counterpart. In
particular, QED$_{3}$ is a superrenormalizable field theory and
therefore free of ultraviolet divergence. Extensive investigations
have found that QED$_{3}$ exhibits a number of interesting physical
properties, such as dynamical chiral symmetry breaking (DCSB)
\cite{Pisarski84, Appelquist86, Appelquist88, Nash89, Atkinson90,
Curtis90, Pennington91, Curtis92, Maris96, Fisher04, Kubota01,
Hands02, Hands04, Gusynin03, Roberts94, Appelquist04, Bashir08,
Bashir09, Braun14}, asymptotic freedom \cite{Appelquist86}, and
permanent confinement \cite{Burden92, Maris95, Bashir08}.
Apparently, QED$_{3}$ is more like four-dimensional QCD than
QED$_{4}$. For this reason, QED$_{3}$ is often considered as a toy
model of QCD$_{4}$ in the context of particle physics. More
interestingly, QED$_{3}$ has proven in the past twenty years to be
an effective low-energy model for several strongly correlated
condensed-matter systems, including high-$T_c$ cuprate
superconductors \cite{Lee06, Affleck88, Ioffe89, Kim97, Kim99,
Rantner01, Rantner02, Franz01, Franz02, Herbut02, Herbut02B, Liu02},
spin-$1/2$ Kagome spin liquid \cite{Ran07, Hermele08}, graphene
\cite{Gusynin04, Gusynin07, Raya08}, quantum critical systems
\cite{Lu14}, and Kane-Mele model with weak-extended Hubbard
interaction \cite{Luo14}.

Although QED$_{3}$ can be well controlled in the ultraviolet regime,
it encounters infrared problems since both Dirac fermions and U(1)
gauge field are massless. For instance, the theory exhibits infrared
divergence if the free gauge boson propagator is utilized in the
perturbative calculations. Appelquist \emph{et al.} showed that such an
infrared divergence can be naturally erased by including dynamical
screening effect of massless fermions into the effective gauge boson
propagator \cite{Appelquist86}. Based on this scheme, Appelquist \emph{et
al.} \cite{Appelquist88} investigated the Dyson-Schwinger equation (DSE) of
fermion self-energy function, and found that the massless fermions
acquire a finite dynamical mass, which induces DCSB, when their
flavor is below certain threshold, $N < N_c$. Most of the existing
analytical and numerical calculations \cite{Nash89, Atkinson90,
Curtis90, Pennington91, Curtis92, Maris96, Fisher04, Kubota01,
Hands02, Hands04, Gusynin03, Roberts94, Appelquist04} agree that the
critical flavor is $N_c \approx 3.5$ at zero temperature.

In the application of QED$_{3}$ to condensed matter systems, DCSB is
usually interpreted as the formation of Heisenberg quantum
antiferromagnetism \cite{Kim99, Liu02, Franz02, Herbut02B,
Herbut02}. It was found that DCSB in QED$_{3}$ with finite gauge
boson mass can still appear if the mass of the gauge boson is not
large enough \cite{Liu03}. This model can describe the coexistence
of antiferromagnetism and superconductivity in high-temperature
cuprate superconductors \cite{Liu03}.

The fate of dynamical mass generation at finite temperature
\cite{Dorey91, Dorey92, Aitchison92, Aitchison94, Lee98,
Triantaphyllou98, Triantaphyllou99, LiLiu10, Feng12A,
Feng12B,Feng12C, Feng13, Yin14, Feng14, Lo14} has also attracted
considerable interest. Notice that finite temperature QED$_{3}$ is
not only interesting from the viewpoint of quantum field theory, but
of practical value since QED$_{3}$ has wide applications in
condensed matter physics. An important problem is to estimate the
critical temperature $T_c$ at which dynamical mass generation is
destroyed by thermal fluctuation.

Unfortunately, the study of dynamical mass generation at finite
temperature seems to be more difficult than the case of zero
temperature. A main obstacle is that the Lorentz invariance is
explicitly broken at finite temperature. The integration over
3-momenta $k = (k_0, k_1, k_2)$ that appears in the DSE is replaced
by an integration over 2-momenta $\mathbf{k} = (k_1, k_2)$ and a
summation over imaginary frequency $k_0 = \omega_{n}$, where
$\omega_{n}$ is the Mastubara frequency with $n$ being an integer.
As a consequence, the DSE becomes much more complicated, so it is
usually necessary to make certain approximations to perform
algebraic calculations. A number of different approximations have
been proposed to study dynamical mass generation in finite
temperature QED$_{3}$ \cite{Dorey91, Dorey92, Aitchison92,
Aitchison94, Lee98, LiLiu10, Feng12A, Feng12B, Feng12C, Feng13,
Yin14, Feng14}. The most frequently used one is the so-called
instantaneous approximation, which assumes that the gauge boson
propagator is completely independent of energy,
\begin{eqnarray}
\Delta_{\mu\nu}(q_{0},\mathbf{q}) &\rightarrow&
\Delta_{\mu\nu}(0,\mathbf{q}).
\end{eqnarray}
In previous works \cite{Dorey91, Dorey92, Aitchison92, Aitchison94,
Feng12A, Feng12B, Feng12C, Feng13, Yin14, Feng14}, the transverse
component of gauge boson propagator, $\Delta_{ij}$, is usually
ignored on top of the instantaneous approximation, i.e.,
\begin{eqnarray}
\Delta_{\mu\nu}(0,\mathbf{q}) \rightarrow\Delta_{00}(0,\mathbf{q}).
\end{eqnarray}
Under these approximations, the DSE for fermion mass becomes much
simpler and can be easily solved. However, as showed in
Ref.~\cite{Lee98}, the transverse component of gauge boson
propagator $\Delta_{ij}$, if incorporated in the DSE, induces an
infrared divergence at finite temperature. To obtain convergent
results, Lee proposed to simply neglect the contribution from
$\Delta_{ij}$ \cite{Lee98}.

However, neglecting the transverse component of gauge boson
propagator is actually problematic. An important reason for
QED$_{3}$ to exhibit dynamical fermion mass generation is that the
gauge interaction is long-ranged, which is guaranteed by the gauge
invariance. If the gauge boson acquires a finite mass, the dynamical
fermion mass generation will be strongly suppressed \cite{Liu03}. At
finite temperature, the Lorentz invariance is explicitly broken,
then the longitudinal and transverse components of gauge boson
propagator behave very differently \cite{Kim97, Kim99, Rantner01,
WangLiu10A, WangLiu10B, JWangLiu12}: the former becomes short-ranged
due to thermal screening, whereas the latter remains long-ranged as
a consequence of gauge invariance. Therefore, the transverse
component of gauge boson propagator should play more significant
role than the longitudinal component at finite temperatures. This
judgement is supported by the extensive recent analysis of
nontrivial properties induced by gauge interaction \cite{Kim97,
Kim99, Rantner01, WangLiu10A, WangLiu10B, JWangLiu12}. Overall, the
transverse component of gauge boson propagator needs to be included
in an appropriate manner in the study of dynamical fermion mass
generation at finite temperature, which is the motivation of our
work.

In this paper, we revisit the issue of dynamical mass generation in
finite temperature QED$_{3}$. In order to obtain physically
meaningful results, we go beyond the widely used instantaneous
approximation, and employ a new approximation that ignores the
energy dependence of dynamical fermion mass, i.e., $m(p_0,
\mathbf{p})\rightarrow m(0,\mathbf{p})$, but maintains both the
longitudinal and transverse components of gauge boson propagator.
Different form the instantaneous approximation, the new
approximation does not completely neglect the energy dependence of
the gauge boson. After performing extensive numerical computations,
we find that the dynamical fermion mass is free of divergence under
the new approximation. We also examine the validity of the new
approximation by comparing to the case of zero temperature, and show
that it is better than the instantaneous approximation.

As aforementioned, QED$_{3}$ can serve as an effective low-energy
field theory for several condensed matter systems \cite{Lee06,
Affleck88, Ioffe89, Kim97, Kim99, Rantner01, Rantner02, Franz01,
Franz02, Herbut02, Herbut02B, Liu02}. In these systems, there are
always certain amount of disorders, which couple to massless
fermions and induce a finite damping rate. Disorders are responsible
to the anomalous behaviors of a plenty of observable quantities of
interacting systems of Dirac fermions \cite{Lee93, Durst00,
Altland02}. Therefore, it is also interesting to study dynamical
mass generation of QED$_{3}$ in the presence of disorders. To
describe the influence of disorders, a finite fermion damping rate
$\Gamma$ is introduced to the fermion propagator
Ref.~\cite{LiLiu10}. As will be shown below, $\Gamma$ plays
analogous role as temperature $T$, so its effect on dynamical mass
generation can be studied in a similar manner to the case of finite
temperature.

The rest of paper is organized as follows. In
Sec.~\ref{Sec:Lagrangian}, we give the Lagrangian and the relevant
propagators. In Sec.~\ref{Sec:GapInstan}, we show that the dynamical
mass is divergent under the instantaneous approximation if
temperature or damping rate is finite. In
Sec.~\ref{Sec:GapNewAppro}, we solve the DSE for dynamical mass by
invoking a new approximation, and find that the results are free of
infrared divergence. The nature of the infrared divergence is also
discussed. In Sec.~\ref{Sec:GapZero}, we examine the validity of the
new approximation by comparing to the case of zero temperature. In
Sec.~\ref{Sec:Conclusion}, we summarize the main results. Detailed
calculations of polarization functions are given in Appendix
\ref{Appendix:Polarization}.

\section{Model and Feynman rules \label{Sec:Lagrangian}}

The Lagrangian density for QED$_{3}$ with $N$ flavors of massless
Dirac fermions is given by
\begin{eqnarray}
\mathcal{L}=\sum_{i=1}^{N}\bar{\psi}_{i}\left(i\partial\!\!\!\slash
+ eA\!\!\!\slash \right)\psi_{i}
-\frac{1}{4}F_{\mu\nu}^{2}.\label{Eq:Langrangian}
\end{eqnarray}
The electromagnetic tensor is related to vector potential $A_{\mu}$
as $F_{\mu\nu} = \partial_{\mu}A_{\nu} -
\partial_{\nu}A_{\mu}$. The fermion is described by a four-component
spinor field $\psi$, whose conjugate spinor field is
$\bar{\psi}=\psi^{\dag}\gamma_{0}$. The $4\times4$ gamma matrixes
are defined as $(\gamma_{0},\gamma_{1},\gamma_{2}) =
(i\sigma_{3},i\sigma_{1},i\sigma_{2})\otimes\sigma_{3}$, which
satisfy the standard Clifford algebra $\{\gamma_{\mu},\gamma_{\nu}\}
= 2g_{\mu\nu}$ with the metric being $g_{\mu\nu} =
\mathrm{diag}(-1,-1,-1)$. In (2+1) dimensions, there are two chiral
matrices $\gamma_{3}$ and $\gamma_{5}$ \cite{Appelquist86,
Bashir05},
\begin{eqnarray}
\gamma_{3}
=i\left(
\begin{array}{cc}
0 & I
\\
-I & I
\end{array}
\right),\qquad
\gamma_{5}
=i\left(
\begin{array}{cc}
0 & I
\\
I & 0
\end{array}
\right),
\end{eqnarray}
which anticommute with $\gamma_{0}$, $\gamma_{1}$ and $\gamma_{2}$.
The Lagrangian shown in Eq.~(\ref{Eq:Langrangian}) respects a
continuous $U(2N)$ chiral symmetry $\psi \rightarrow e^{i\theta
\gamma_{3,5}}\psi$, where $\theta$ is an arbitrary constant. Once a
finite fermion mass is dynamically generated, the global $U(2N)$
chiral symmetry is spontaneously broken down to its subgroup
$U(N)\times U(N)$. In this paper, we consider a general large $N$
and perform perturbative expansion in powers of $1/N$. For
convenience, we work in units with $\hbar=k_{B}=1$ and set the
velocity $v_{F} \equiv 1$.

In the Euclidian space, the free propagator of massless fermions is
\begin{eqnarray}
G_{0}(k) = \frac{1}{k_{\mu}\gamma_{\mu}}
\end{eqnarray}
at zero temperature. The free propagator of gauge boson can be
written as
\begin{eqnarray}
\Delta_{\mu\nu}^{(0)}(q) = \frac{1}{q^{2}}\left(\delta_{\mu\nu} -
\frac{q_{\mu}q_{\nu}}{q^{2}}\right)
\end{eqnarray}
in the Landau gauge. After including the dynamical screening effect
due to fermions, the effective gauge boson propagator takes the form
\begin{eqnarray}
\Delta_{\mu\nu}(q) = \frac{1}{q^2+\Pi(q)}\left(\delta_{\mu\nu} -
\frac{q_{\mu}q_{\nu}}{q^{2}}\right).
\end{eqnarray}
To the lowest order of $1/N$-expansion, the polarization function
can be obtained from the polarization tensor through the
relationship
\begin{eqnarray}
\Pi_{\mu\nu}(q) &=& Ne^{2}\int\frac{d^3k}{(2\pi)^{3}}\mathrm{Tr}
\left[G_{0}(k)\gamma_{\mu}G_{0}(k+q)\gamma_{\nu}\right]\nonumber \\
&=&\Pi(q)\left(\delta_{\mu\nu}-\frac{q_{\mu}q_{\nu}}{q^{2}}\right),
\end{eqnarray}
where $\Pi(q)=\alpha q$ with $\alpha=\frac{Ne^{2}}{8}$
\cite{Appelquist88}.

Pisarski \cite{Pisarski84} first carefully studied DCSB in QED$_3$
by means of non-perturbative DSEapproach, and showed that DCSB takes
place for any finite flavor $N$. However, subsequent analysis of
Appelquist \emph{et. al.} \cite{Appelquist88} found that DCSB can
occur only when the fermion flavor $N$ is smaller than a critical
value $N_c$, which is $N_c = 32/\pi^2$ to the lowest order of
$1/N$-expansion. Nash \cite{Nash89} then examined the effect of
next-to-leading order correction and obtained a critical flavor $N_c
= (4/3)32/\pi^2$. Pennington \emph{et. al.} \cite{Pennington91,
Curtis92} included the wave renormalization function and claimed
that DCSB can be realized for any flavor $N$, although the
corresponding dynamical fermion mass decreases exponentially with
increasing $N$. Nevertheless, their analysis ignored the influence
of wave renormalization function on the polarization. Later, Maris
\cite{Maris96} studied DCSB by solving a set of self-consistent DSEs
for fermion propagator and polarization, which contain vertex
corrections and therefore satisfy the Ward-Takahashi (WT)
identities. The calculations of Ref.~\cite{Maris96} arrived at a
finite critical flavor $N_{c} \approx 3.3$, which is close to that
of Ref.~\cite{Appelquist88}. The key difference between the
treatments of Pennington \emph{et al}. \cite{Pennington91} and Maris
\cite{Maris96} is that the latter included the interaction
correction to the polarization whereas the former did not. It turns
out that an appropriate approximation plays a crucial role in the
DSE analysis of DCSB. More refined calculations of Fisher \emph{et
al}. revealed a finite critical flavor $N_{c} \approx 4$
\cite{Fisher04}. The gauge invariance of $N_{c}$ is also discussed
\cite{Nash89, Bashir08, Bashir09}. Apart from the DSE approach, this
problem can be studied by renormalization group method, which found
that $3 < N_c < 4$ \cite{Kubota01}. The critical flavor $N_{c}$
obtained in lattice Monto Carlo simulations \cite{Hands02, Hands04}
is much smaller than that obtained by means of DSEs. However,
Gusynin \emph{et al.} argued that the difference is attributed to
the finite volume effect introduced in lattice simulations
\cite{Gusynin03}. It is fairy to say that, although there is still
some debate \cite{Appelquist04, Bashir08, Bashir09}, most studies
have obtained a finite critical flavor for DCSB in QED$_3$, which is
roughly $N_c \approx 3.5$ at $T=0$.

As we go to finite temperature, the dynamical fermion mass is
expected to be rapidly suppressed by thermal fluctuation. The
temperature scale $T_c$ at which the dynamical mass vanishes defines
the critical temperature. At $T \neq 0$, we write the fermion
propagator in the standard Matsubara formalism as
\begin{eqnarray}
G(k_{0},\mathbf{k}) = \frac{1}{(k_{0} +
\Gamma\mathrm{sgn}\left(k_{0}\right))\gamma_{0} +
\mathbf{\gamma}\cdot\mathbf{k} + m_{0}},\label{Eq:GTGammaMass}
\end{eqnarray}
where $k_{0}=(2n+1)\pi T$ with $n$ being an integer. Here, we
introduce a constant $\Gamma$ to represent the fermion damping rate
generated by disorder scattering. This quantity measures the
strength of the fermion damping effect. For more explanation of the
origin and the physical effect of the constant $\Gamma$, please see
Ref.~\cite{LiLiu10}. To the leading order of $1/N$-expansion, the
polarization tensor is given by
\begin{eqnarray}
\Pi_{\mu\nu}(q_{0},\mathbf{q},T,m_0,\Gamma) &=& \frac{Ne^{2}}{\beta}
\sum_{n=-\infty}^{+\infty}\int\frac{d^{2}\mathbf{k}}{(2\pi)^{2}}
\mathrm{Tr}\left[G(k_{0},\mathbf{k})\right.\nonumber \\
&&\times\left.\gamma_{\mu}G(k_{0} + q_{0},\mathbf{k} +
\mathbf{q})\gamma_{\nu}\right],
\end{eqnarray}
where $q_0 = 2n'\pi T$ with $n'$ being an integer and $\beta =
\frac{1}{T}$. The effective propagator of gauge boson now becomes
\begin{eqnarray}
\Delta_{\mu\nu}(q_{0},\mathbf{q}) &=& \frac{A_{\mu\nu}}{q_{0}^2 +
\mathbf{q}^{2}+\Pi_{A}(q_{0},\mathbf{q})} \nonumber \\
&& +\frac{B_{\mu\nu}}{q_{0}^{2} +
\mathbf{q}^2+\Pi_{B}(q_{0},\mathbf{q})},
\end{eqnarray}
where
\begin{eqnarray}
A_{\mu\nu} &=& \left(\delta_{\mu0} - \frac{q_{\mu}q_{0}}{q^2}\right)
\frac{q^2}{\mathbf{q}^2}\left(\delta_{0\nu} -
\frac{q_{0}q_{\nu}}{q^2}\right), \\
B_{\mu\nu} &=& \delta_{\mu i}\left(\delta_{ij} -
\frac{q_{i}q_{j}}{\mathbf{q}^{2}}\right)\delta_{j\nu}.
\end{eqnarray}
$A_{\mu\nu}$ and $B_{\mu\nu}$ are orthogonal and satisfy
\begin{eqnarray}
A_{\mu\nu}+B_{\mu\nu}=\delta_{\mu\nu}-\frac{q_{\mu}q_{\nu}}{q^2}.
\end{eqnarray}
The functions $\Pi_{A}$ and $\Pi_{B}$ are defined as
\begin{eqnarray}
\Pi_{A}=\frac{q^2}{\mathbf{q}^2}\Pi_{00},\qquad\Pi_{B} =
\Pi_{ii}-\frac{q_{0}^{2}}{\mathbf{q}^2}\Pi_{00}.
\end{eqnarray}
The calculational details of the relevant polarization functions are
shown in Appendix \ref{Appendix:Polarization}.

In the following, we will consider two approximations, namely the
popular instantaneous approximation and a new approximation to be
explained below. We neglect the energy-dependence of the
polarization functions and also the feedback of dynamical mass to
the polarization functions. Firstly, in the limit that $q_{0}=0$,
$\Gamma=0$, and $m_{0}=0$, the polarization functions are
\begin{eqnarray}
\Pi_{A}(\mathbf{q},T) &=& \frac{16\alpha T}{\pi}\int_{0}^{1}dx
\ln\left[2\cosh\left(\frac{\sqrt{x(1-x)\mathbf{q}^2}}{2T}\right)\right],
\label{Eq:PiAT} \nonumber \\
\\
\Pi_{B}(\mathbf{q},T) &=& \frac{8\alpha}{\pi}
\int_{0}^{1}dx\sqrt{x(1-x)\mathbf{q}^2} \nonumber \\
&& \times \tanh\left(\frac{\sqrt{x(1-x)\mathbf{q}^2}}{2T}\right).
\label{Eq:PiBT}
\end{eqnarray}
We have used $\alpha=\frac{Ne^2}{8}$. Secondly, in the limit that
$q_{0}=0$, $T=0$ and $m_{0}=0$, the polarization functions are
\begin{eqnarray}
\Pi_{A}(\mathbf{q},\Gamma) &=& \frac{16\alpha}{\pi^2}
\left\{\Gamma\ln\left(\frac{\Lambda}{\Gamma}\right) +\Gamma
\left[1+\frac{\sqrt{\mathbf{q}^2 +
4\Gamma^2}}{2|\mathbf{q}|}\right.\right.\nonumber
\\
&&\left.\times\ln\left(\frac{\sqrt{\mathbf{q}^2 + 4\Gamma^{2}} -
|\mathbf{q}|} {\sqrt{\mathbf{q}^2 +
4\Gamma^{2}}+|\mathbf{q}|}\right)\right] \nonumber \\
&& + \int_{0}^{1}dx\sqrt{x(1-x)\mathbf{q}^{2}}\nonumber \\
&&\left.\times\arctan \left(\frac{\sqrt{x(1 -
x)\mathbf{q}^{2}}}{\Gamma}\right)\right\},\label{Eq:PiAGamma}
\\
\Pi_{B}(\mathbf{q},\Gamma) &=& \frac{16\alpha}{\pi^2}
\int_{0}^{1}dx\sqrt{x(1-x)\mathbf{q}^{2}} \nonumber \\
&&\times \arctan
\left(\frac{\sqrt{x(1-x)\mathbf{q}^{2}}}{\Gamma}\right).
\label{Eq:PiBGamma}
\end{eqnarray}
From these expressions, we can see that the fermion damping rate
$\Gamma$ plays a very similar role to temperature $T$, which allows
us to study the effects of temperature and fermion damping using the
same scheme. To simplify later calculations, we can further
approximate Eqs.~(\ref{Eq:PiAT}) and (\ref{Eq:PiBT}) by
\cite{Aitchison92}
\begin{eqnarray}
\Pi_{A}(\mathbf{q},T)&\approx&\alpha\left[|\mathbf{q}| +
c_{1}T\exp\left(-\frac{|\mathbf{q}|}{c_{1}T}\right)\right],
\label{Eq:PiATAp}
\\
\Pi_{B}(\mathbf{q},T) &\approx&\alpha|\mathbf{q}|\tanh\left(
\frac{c_{2}|\mathbf{q}|}{T}\right), \label{Eq:PiBTAp}
\end{eqnarray}
where $c_{1} = 16\ln2/\pi$ and $c_2 = 2/3\pi$. Analogously,
Eqs.~(\ref{Eq:PiAGamma}) and (\ref{Eq:PiBGamma}) can be approximated
by
\begin{eqnarray}
\Pi_{A}(\mathbf{q},\Gamma)&\approx&\frac{16\alpha\Gamma}{\pi^2}
\ln\left(\frac{\Lambda}{\Gamma}\right)
+\frac{2\alpha|\mathbf{q}|}{\pi}
\arctan\left(\frac{c_{3}|\mathbf{q}|}{\Gamma}\right), \nonumber\\
\label{Eq:PiAGammaAp} \\ \Pi_{B}(\mathbf{q},\Gamma) &\approx&
\frac{2\alpha|\mathbf{q}|}{\pi}\arctan\left(
\frac{c_{3}|\mathbf{q}|}{\Gamma}\right), \label{Eq:PiBGammaAp}
\end{eqnarray}
where $c_3 = 4/3\pi$. It can be checked numerically that expressions
Eqs.~(\ref{Eq:PiATAp})-(\ref{Eq:PiBGammaAp}) are very good
approximations for both the high- and low-momentum behaviors of
Eqs.~(\ref{Eq:PiAT})-(\ref{Eq:PiBGamma}). In the following sections,
we will use Eqs.~(\ref{Eq:PiATAp})-(\ref{Eq:PiBGammaAp}) to analyze
and numerically solve the DSEs for dynamical fermion mass at finite
temperature and/or finite fermion damping rate.

\section{DSE under instantaneous approximation\label{Sec:GapInstan}}

In this section, we present the DSE for dynamical mass under
the widely used instantaneous approximation. We will show that the
solutions of DSE are divergent in the infrared region
whenever $T \neq 0$ and therefore ill-defined. Such an infrared
divergence also exists if $T=0$ but $\Gamma \neq 0$.

The free fermion propagator is
\begin{eqnarray}
G_{0}(k_{0},\mathbf{k}) = \frac{1}{(k_{0} + \Gamma\mathrm{sgn}
\left(k_{0}\right))\gamma_{0}+\mathbf{\gamma}\cdot\mathbf{k}}.
\label{Eq:FermionPropagatorFree}
\end{eqnarray}
Due to gauge interaction, the fermion may become massive and the
propagator is renormalized to
\begin{widetext}
\begin{eqnarray}
G(k_{0},\mathbf{k}) = \frac{1}{(k_{0} +
\Gamma\mathrm{sgn}\left(k_{0}\right))\gamma_{0} +
\mathbf{\gamma}\cdot\mathbf{k} +m(k_{0},\mathbf{k},T,\Gamma)}.
\nonumber \\
\label{Eq:FermionPropagatorFull}
\end{eqnarray}
Here, to the lowest order of $1/N$-expansion, we neglect the wave
renormalization function. Now, $G_{0}(k_{0},\mathbf{k})$ is related
to $G(k_{0},\mathbf{k})$ through the following DSE,
\begin{eqnarray}
G^{-1}(p_{0},\mathbf{p}) = G_{0}^{-1}(p_{0},\mathbf{p}) +
m(p_{0},\mathbf{p}), \label{Eq:DysonEquation}
\end{eqnarray}
where
\begin{eqnarray}
m(p_{0},\mathbf{p}) = \frac{e^{2}}{\beta}
\sum_{k_{0}}\int\frac{d^2\mathbf{k}}{(2\pi)^2}
\gamma_{\mu}G_{F}(k_{0},\mathbf{k})\gamma_{\nu}
\Delta_{\mu\nu}(q_{0},\mathbf{q})\label{Eq:SelfEnergySC}
\end{eqnarray}
with $q_0 = p_0 - k_0$ and $\mathbf{q} = \mathbf{p} - \mathbf{k}$.
Substituting Eqs.~(\ref{Eq:FermionPropagatorFree}) and
(\ref{Eq:FermionPropagatorFull}) into Eq.~(\ref{Eq:SelfEnergySC}),
and then giving an explicit $T$- and $\Gamma$-dependence to the
fermion mass, we obtain
\begin{eqnarray}
m(p_{0},\mathbf{p},T,\Gamma) &=& \frac{e^{2}}{\beta}\sum_{k_{0}}
\int \frac{d^2\mathbf{k}}{(2\pi)^2}
\frac{m(k_{0},\mathbf{k},T,\Gamma)}{(k_{0} +
\Gamma\mathrm{sgn}\left(k_{0}\right))^{2}+\mathbf{k}^{2}
+m^2(k_{0},\mathbf{k},T,\Gamma)}\nonumber \\
&&\times\left[\frac{1}{q_{0}^2 + \mathbf{q}^{2} +
\Pi_{A}(q_{0},\mathbf{q},T,\Gamma)}+\frac{1}{q_{0}^{2}+\mathbf{q}^2 +
\Pi_{B}(q_{0},\mathbf{q},T,\Gamma)}\right].\label{Eq:GapEqOriginal}
\end{eqnarray}
In previous studies \cite{Dorey91, Dorey92, Aitchison92,
Aitchison94, Feng12A,Feng12B, Feng12C, Feng13, Yin14, Feng14}, the
key assumption beneath the instantaneous approximation is to
completely ignore the energy-dependence of gauge boson propagator.
Apart from this approximation, the transverse component of gauge
boson propagator is widely neglected, i.e.,
\begin{eqnarray}
\Delta_{\mu\nu}(q_{0},\mathbf{q}) \rightarrow
\Delta_{00}(0,\mathbf{q}).
\end{eqnarray}
Now we follow the instantaneous approximation and drop the energy
dependence of gauge boson propagator. Nevertheless, as illustrated
in recent works on the nontrivial properties of QED$_{3}$
\cite{Kim97, Kim99, Rantner01, WangLiu10A, WangLiu10B, JWangLiu12},
the transverse component of gauge interaction plays more important
role than the longitudinal component at finite temperature, since
the latter becomes short-ranged after acquiring an effective thermal
mass proportional to $T$. If $T = 0$ and $\Gamma \neq 0$, the
longitudinal part of gauge interaction also becomes short-ranged due
to static screening caused by disorder scattering. In any case, the
gauge invariance ensures that the transverse component of gauge
interaction is strictly long-ranged. It is therefore not appropriate
to neglect the transverse component. For completeness, here we keep
them both and write the gauge boson propagator as
\begin{eqnarray}
\Delta_{\mu\nu}(0,\mathbf{q}) = \frac{\delta_{\mu0}\delta_{0\nu}}
{\mathbf{q}^2 + \Pi_{A}(0,\mathbf{q})} + \frac{\delta_{\mu
i}\left(\delta_{ij} -
\frac{q_{i}q_{j}}{\mathbf{q}^{2}}\right)\delta_{j\nu}}{\mathbf{q}^2
+ \Pi_{B}(0,\mathbf{q})}.
\end{eqnarray}
The corresponding DSE becomes
\begin{eqnarray}
m(\mathbf{p},T,\Gamma) = \frac{e^{2}}{\beta}\sum_{k_{0}}\int
\frac{d^2\mathbf{k}}{(2\pi)^2}\frac{m(\mathbf{k},T,\Gamma)}{(k_{0} +
\Gamma\mathrm{sgn}(k_{0}))^{2} + \mathbf{k}^{2} +
m^2(\mathbf{k},T,\Gamma)}
\left[\frac{1}{\mathbf{q}^{2}+\Pi_{A}(0,\mathbf{q},T,\Gamma)} +
\frac{1}{\mathbf{q}^2 + \Pi_{B}(0,\mathbf{q},T,\Gamma)}\right].
\end{eqnarray}

Once the instantaneous approximation is adopted, the dynamical mass
$m$ completely loses its dependence on energy. Therefore, the
frequency summation in the DSE can be written as
\begin{eqnarray}
S_{1} = \sum_{k_{0}}\frac{1}{\left[\left(k_{0} +
\Gamma\mathrm{sgn}\left(k_{0}\right)\right)^{2}+\mathbf{k}^{2} +
m^2(\mathbf{k},T,\Gamma)\right]} =
\left(\frac{\beta}{2\pi}\right)^{2}\sum_{n=-\infty}^{\infty}
\frac{1}{\left[\left(n+\frac{1}{2} + X
\mathrm{sgn}\left(n+\frac{1}{2}\right)\right)^{2} + Y^2\right]},
\end{eqnarray}
where $X = \frac{\beta}{2\pi}\Gamma$, and $Y =
\frac{\beta}{2\pi}\sqrt{\mathbf{k}^{2}+m^2(\mathbf{k},T,\Gamma)}$.
With the help of an identity
\begin{eqnarray}
S(X,Y) = \sum_{n=0}^{\infty}\frac{1}{(n+X)^{2}+Y^{2}} =
\frac{1}{2Yi}\left[\psi(X+iY)-\psi(X-iY)\right]
\end{eqnarray}
where $\psi\left(z\right)$ is Digamma function, we get
\begin{eqnarray}
S_{1} = \frac{\beta^2}{2\pi^2}\sum_{n=0}^{\infty}
\frac{1}{\left[\left(n+\frac{1}{2} + X\right)^{2} + Y^2\right]} =
\frac{\beta^2}{2\pi^2Y}
\mathrm{Im}\left[\psi\left(\frac{1}{2}+X+iY\right)\right].
\end{eqnarray}
Carrying out frequency summation leads to
\begin{eqnarray}
m(\mathbf{p},T,\Gamma) &=& \frac{8\alpha}{N}\int
\frac{d^2\mathbf{k}}{(2\pi)^2}
\frac{m(\mathbf{k},T,\Gamma)}{\sqrt{\mathbf{k}^{2} +
m^2(\mathbf{k},T,\Gamma)}} \frac{1}{\pi}
\mathrm{Im}\left[\psi\left(\frac{1}{2}+\frac{\Gamma}{2\pi T} +
i\frac{\sqrt{\mathbf{k}^{2} + m^2(\mathbf{k},T,\Gamma)}}{2\pi
T}\right)\right]\nonumber \\
&&\times\left[\frac{1}{\mathbf{q}^{2}+\Pi_{A}(0,\mathbf{q},T,\Gamma)}
+\frac{1}{\mathbf{q}^2+\Pi_{B}(0,\mathbf{q},T,\Gamma)}\right].
\end{eqnarray}

This equation can be further simplified. First, we assume $\Gamma=0$
and $T \neq 0$, and then rewrite the DSE as
\begin{eqnarray}
m(\mathbf{p},T) = \frac{4\alpha}{N}\int
\frac{d^2\mathbf{k}}{(2\pi)^2}
\frac{m(\mathbf{k},T)}{\sqrt{\mathbf{k}^{2} +
m^2(\mathbf{k},T)}}\tanh\left(\frac{\sqrt{\mathbf{k}^{2} +
m^2(\mathbf{k},T)}}{2T}\right) \left[\frac{1}{\mathbf{q}^{2} +
\Pi_{A}(\mathbf{q},T)} + \frac{1}{\mathbf{q}^2 +
\Pi_{B}(\mathbf{q},T)}\right],\label{Eq:GapFTInstanAppro}
\end{eqnarray}
where $\mathrm{Im}\left[\psi\left(\frac{1}{2}+\frac{i\xi}{2\pi
T}\right)\right] = \frac{\pi}{2}\tanh\left(\frac{\xi}{2T}\right)$ is
used in the derivation. Secondly, at $T=0$ and $\Gamma \neq 0$, we
have
\begin{eqnarray}
m(\mathbf{p},\Gamma) = \frac{8\alpha}{N}\int
\frac{d^2\mathbf{k}}{(2\pi)^2}
\frac{m(\mathbf{k},\Gamma)}{\sqrt{\mathbf{k}^{2} +
m^2(\mathbf{k},\Gamma)}} \frac{1}{\pi}
\arctan\left(\frac{\sqrt{\mathbf{k}^{2} +
m^2(\mathbf{k},\Gamma)}}{\Gamma}\right)
\left[\frac{1}{\mathbf{q}^{2}+\Pi_{A}(\mathbf{q},\Gamma)} +
\frac{1}{\mathbf{q}^2+\Pi_{B}(\mathbf{q},\Gamma)}\right],
\label{Eq:GapFGammaInstanAppro}
\end{eqnarray}
where $\lim_{T \rightarrow 0}
\mathrm{Im}\left[\psi\left(\frac{1}{2}+\frac{\Gamma+i\xi}{2\pi
T}\right)\right] = \arctan\left(\frac{\xi}{\Gamma}\right)$. The
similarity between the above two equations can be clearly seen. At
$T = 0$, the constant fermion damping rate $\Gamma$ can be
considered as certain effective temperature. This is an important
reason for us to discuss these two cases simultaneously in this
paper.

\begin{figure}[htbp]
\includegraphics[width=3in]{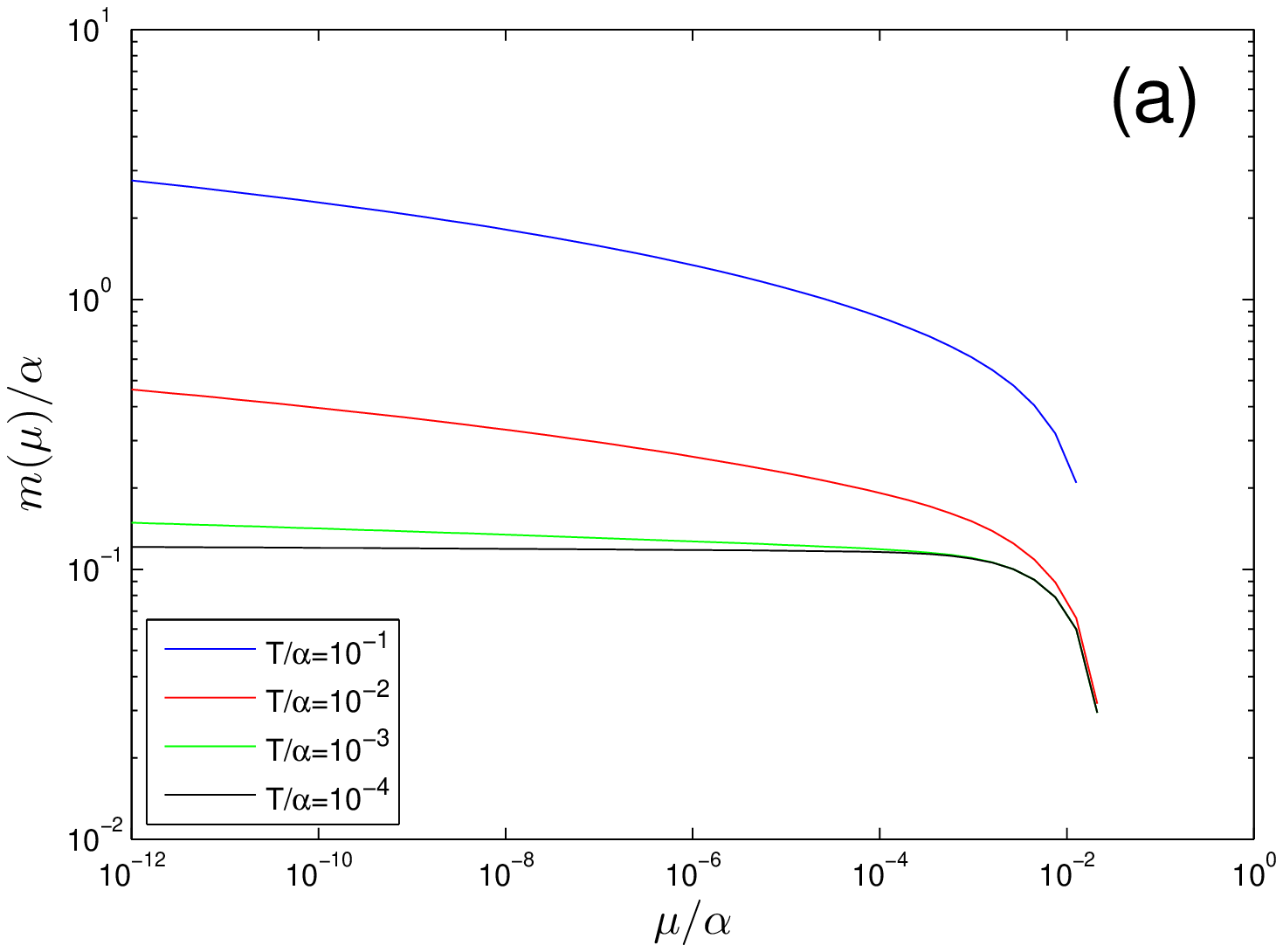}
\includegraphics[width=3in]{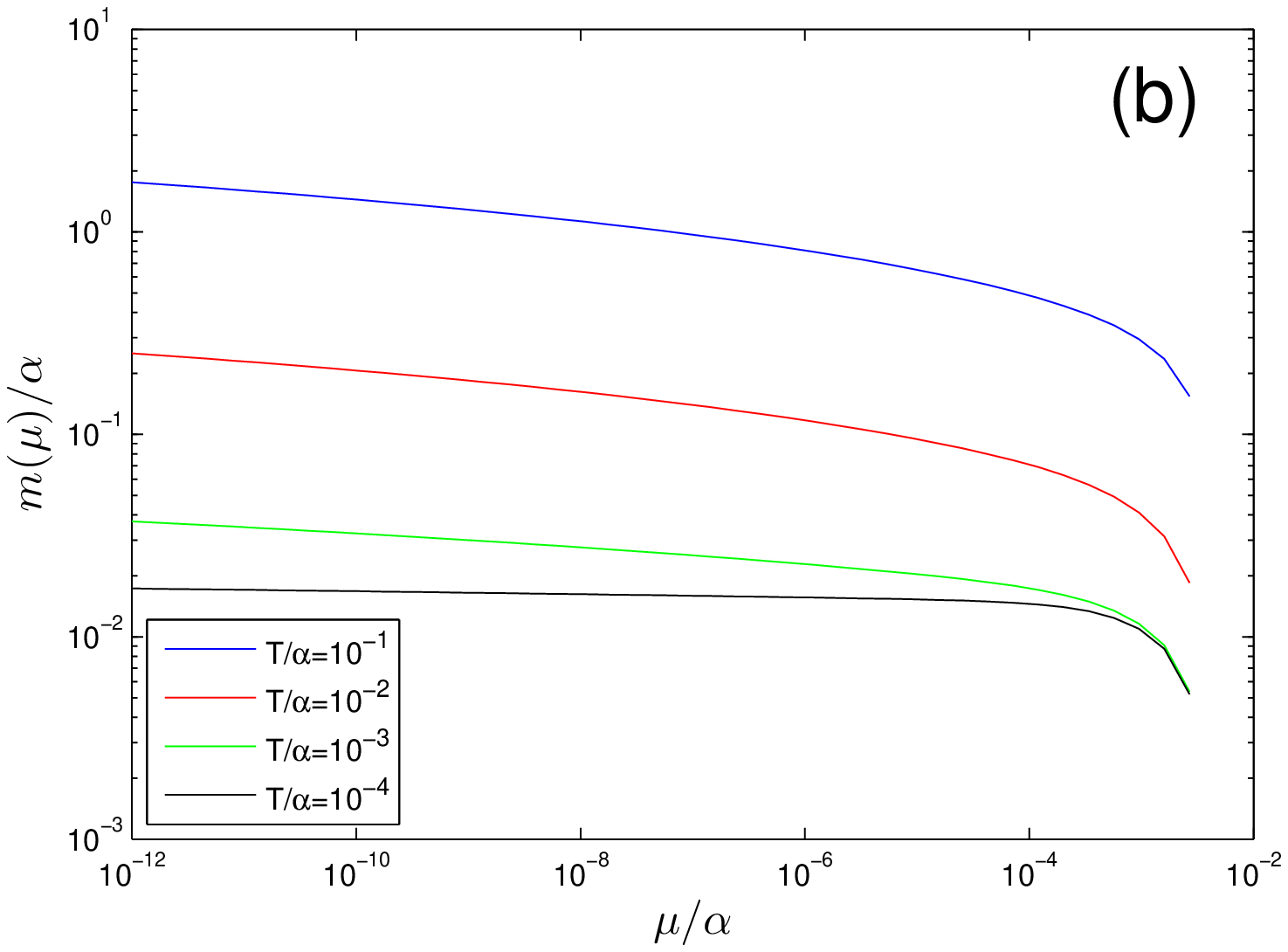}
\caption{Relationship between $m(\mu)/\alpha$ and $\mu/\alpha$ for
different temperatures at $\Gamma=0$ with $N=2,3$ in (a) and (b).}
\label{Fig:GapMuFTInstan}
\end{figure}
\begin{figure}[htbp]
\includegraphics[width=3in]{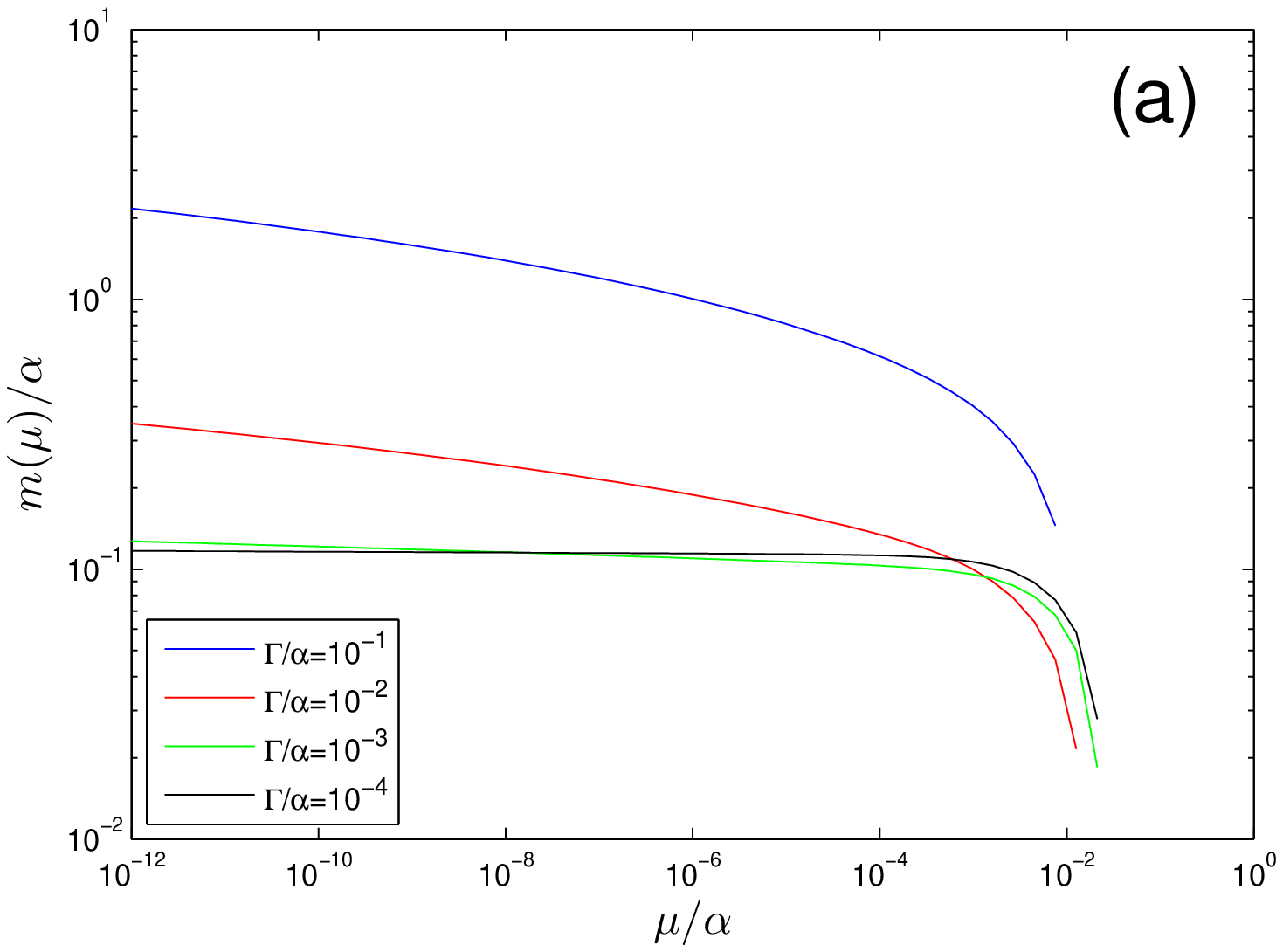}
\includegraphics[width=3in]{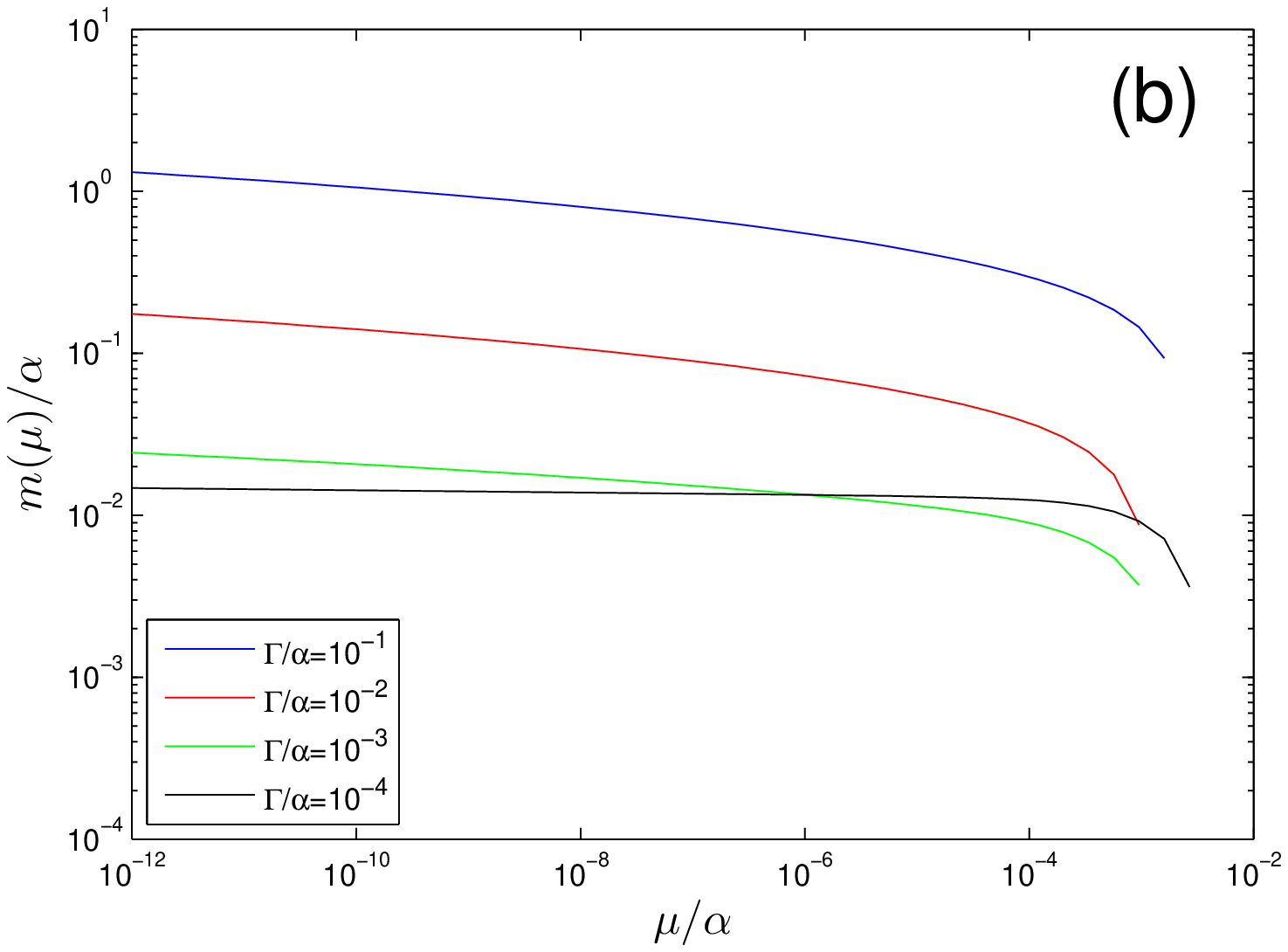}
\caption{Relationship between $m(\mu)/\alpha$ and $\mu/\alpha$ for
different $\Gamma$ at $T=0$ with $N=2,3$ in (a) and (b)
respectively.} \label{Fig:GapMuFGammaInstan}
\end{figure}

To solve the DSE, it is convenient to introduce an infrared cutoff
$\mu$. This amounts to assume that the system has a finite volume
\cite{Gusynin03, Liu09}. For finite $\mu$, the dynamical fermion
mass obtained from DSE is always free of infrared divergence. Our
aim is to examine whether the dynamical mass is still well defined
as $\mu \rightarrow 0$. We employ an infrared cutoff $\mu_{1}$ for
momentum $|\mathbf{k}|$ and replace $|\mathbf{q}|$ with
$|\mathbf{q}|+\mu_{2}$. Usually, $\mu_{1}$ and $\mu_{2}$ should
satisfy $\mu_{1}\sim \mu_{2}$. For simplicity, we assume $\mu_{1} =
\mu_{2} = \mu$. Now the DSE at $\Gamma=0$ and $T \neq 0$ becomes
\begin{eqnarray}
m(\mathbf{p},T) &=& \frac{\alpha}{N\pi^2}\int_{\mu}^{\Lambda}
d|\mathbf{k}||\mathbf{k}| \int_{0}^{2\pi}d\varphi
\frac{m(\mathbf{k},T)}{\sqrt{\mathbf{k}^{2} +
m^2(\mathbf{k},T)}}\tanh\left(\frac{\sqrt{\mathbf{k}^{2} +
m^2(\mathbf{k},T)}}{2T}\right) \nonumber \\
&&\times \left[\frac{1}{\left(|\mathbf{q}|+\mu\right)^{2} +
\Pi_{A}(|\mathbf{q}|+\mu,T)} +
\frac{1}{\left(|\mathbf{q}|+\mu\right)^2 +
\Pi_{B}(|\mathbf{q}|+\mu,T)}\right],
\end{eqnarray}
where $\varphi$ is the angle between $\mathbf{k}$ and $\mathbf{p}$.
As usual, we choose $\Lambda=\alpha$\cite{Appelquist88}. The DSE at
$T=0$ and $\Gamma \neq 0$ can be transformed analogously.

For different finite temperatures, the dynamical mass at the lowest
momentum $\mu$ with $N=2$ and $N=3$ are shown in
Figs.~\ref{Fig:GapMuFTInstan} (a) and (b) respectively. As $\mu
\rightarrow 0$, $m(\mu)$ does not converge to a finite value, but is
divergent. As $T$ further decreases, $m(\mu)$ grows more slowly, but
still divergent at finite $T$. For different values of $\Gamma$, the
relations between $m(\mu)$ and $\mu$ with $N=2$ and $N=3$ are
displayed in Fig.~\ref{Fig:GapMuFGammaInstan}(a) and (b)
respectively. Numerical calculations show that $m(\mu)$ is ill
defined as $\mu\rightarrow$ whenever $\Gamma \neq 0$. Apparently,
the instantaneous approximation leads to divergent results.
\end{widetext}

\section{Dynamical mass generation under a new approximation\label{Sec:GapNewAppro}}

We have shown in the last section that the widely used instantaneous
approximation leads to infrared divergence in the dynamical fermion
mass obtained at finite temperature or in the presence of a finite
fermion damping rate, when both the longitudinal and transverse
components of gauge boson propagator are adopted. It is therefore
necessary to go beyond this approximation and seek a new
approximation that could yield convergent results. We find it very
helpful to employ a treatment utilized by Ref.~\cite{Gamayun10} in
the study of excitonic insulating transition in graphene
\cite{Khveshchenko01, Gorbar02, Liu09, Gamayun10, WangLiu12}. The
key assumption of this treatment is to neglect the energy dependence
of dynamical fermion mass, which was shown \cite{Gamayun10} to
generate physically reliable dynamical mass for Dirac fermions.

Let us start from the DSE given by (28). Now we assume the
dynamical fermion mass does not depend on energy, namely
\begin{eqnarray}
m(p_{0},\mathbf{p},T,\Gamma)\rightarrow m(\mathbf{p},T,\Gamma).
\label{Eq:NewApproStepA}
\end{eqnarray}
After making this approximation, the DSE
(\ref{Eq:GapEqOriginal}) can be written in the following form
\begin{eqnarray}
m(\mathbf{p},T,\Gamma) &=& \frac{8\alpha}{N\beta}\sum_{k_{0}}
\int\frac{d^2\mathbf{k}}{(2\pi)^2} \nonumber \\
&& \times \frac{m(\mathbf{k},T,\Gamma)}{(k_{0} +
\Gamma\mathrm{sgn}\left(k_{0}\right))^{2} + \mathbf{k}^{2} +
m^2(\mathbf{k},T,\Gamma)}\nonumber \\
&&\times \left[\frac{1}{k_{0}^2 +
\mathbf{q}^{2}+\Pi_{A}(k_{0},\mathbf{q},T,\Gamma)}\right.
\nonumber \\
&&\left.+\frac{1}{k_{0}^{2} + \mathbf{q}^2 +
\Pi_{B}(k_{0},\mathbf{q},T,\Gamma)}\right].
\label{Eq:GapEqFTFGammaNewApprA}
\end{eqnarray}
Notice the effective gauge boson propagator still retains an
explicit energy dependence, which is thus different from the
instantaneous approximation. However, the summation over $k_0$
cannot be performed exactly due to the complicated $k_0$ dependence
of polarization functions. To further simplify the DSE, we drop the
energy dependence of polarization functions, i.e.,
\begin{eqnarray}
\Pi_{A}(k_{0},\mathbf{q},T,\Gamma)\rightarrow
\Pi_{A}(0,\mathbf{q},T,\Gamma),\label{Eq:NewApproStepB1}
\\
\Pi_{B}(k_{0},\mathbf{q},T,\Gamma)\rightarrow
\Pi_{B}(0,\mathbf{q},T,\Gamma).\label{Eq:NewApproStepB2}
\end{eqnarray}
The DSEs then can be formally written as
\begin{eqnarray}
m(\mathbf{p},T) &=& \frac{8\alpha}{N\beta}\sum_{k_{0}}\int
\frac{d^2\mathbf{k}}{(2\pi)^2} \frac{m(\mathbf{k},T)}{k_{0}^{2} +
\mathbf{k}^{2} +
m^2(\mathbf{k},T)}\nonumber \\
&&\times \left[\frac{1}{k_{0}^2 +
\mathbf{q}^{2}+\Pi_{A}(\mathbf{q},T)}\right.\nonumber\\
&&\left.+\frac{1}{k_{0}^{2} + \mathbf{q}^2 +
\Pi_{B}(\mathbf{q},T)}\right]\label{Eq:GapEqFTNewApprB}
\end{eqnarray}
in the case of $T \neq 0$ and $\Gamma = 0$, and to
\begin{eqnarray}
m(\mathbf{p},\Gamma) &=& \frac{8\alpha}{N}\int\frac{dk_{0}}{2\pi}
\int \frac{d^2\mathbf{k}}{(2\pi)^2} \nonumber \\
&&\times\frac{m(\mathbf{k},\Gamma)}{(k_{0} +
\Gamma\mathrm{sgn}\left(k_{0}\right))^{2} + \mathbf{k}^{2} +
m^2(\mathbf{k},\Gamma)}\nonumber \\
&&\times \left[\frac{1}{k_{0}^2 +
\mathbf{q}^{2}+\Pi_{A}(\mathbf{q},\Gamma)}\right.
\nonumber \\
&&\left.+\frac{1}{k_{0}^{2} + \mathbf{q}^2 +
\Pi_{B}(\mathbf{q},\Gamma)}\right]
\label{Eq:GapEqFGammaNewApprB}
\end{eqnarray}
in the case of $\Gamma \neq 0$ and $T=0$. The infrared behaviors of
these two equations will be further analyzed later. It is now
straightforward to sum over $k_0$, which leads us to the following
expressions:
\begin{widetext}
\begin{eqnarray}
m(\mathbf{p},T) &=&
\frac{4\alpha}{N}\int\frac{d^2\mathbf{k}}{(2\pi)^2}
\left\{\frac{m(\mathbf{k},T)}{\mathbf{q}^{2} +
\Pi_{A}(|\mathbf{q}|,T) - \mathbf{k}^{2} - m^2(\mathbf{k},T)}\right.
\nonumber \\
&&\times \left[\frac{1}{\sqrt{\mathbf{k}^{2} +
m^2(\mathbf{k},T)}}\tanh\left(\frac{\sqrt{\mathbf{k}^{2} +
m^2(\mathbf{k},T)}}{2 T}\right)
-\frac{1}{\sqrt{\mathbf{q}^{2} + \Pi_{A}(\mathbf{q},T)}}
\tanh\left(\frac{\sqrt{\mathbf{q}^{2}+\Pi_{A}(\mathbf{q},T)}}{2
T}\right)\right] \nonumber \\
&&\left.+\left[\Pi_{A}(\mathbf{q},T) \rightarrow
\Pi_{B}(\mathbf{q},T)\right]\right\},\\ \label{Eq:GapEqFTNewAppr}
m(\mathbf{p},\Gamma) &=& \frac{8\alpha}{N\pi}\int
\frac{d^2\mathbf{k}}{(2\pi)^2}
\left\{\frac{m(\mathbf{k},\Gamma)}{\left(\Gamma^2+\mathbf{k}^{2} +
m^2(\mathbf{k},\Gamma)-\mathbf{q}^{2} -
\Pi_{A}(\mathbf{q},\Gamma)\right)^2 + 4\Gamma^2 \left(\mathbf{q}^{2}
+ \Pi_{A}(\mathbf{q},\Gamma)\right)}\right.\nonumber \\
&&\times\left[\frac{\Gamma^2-\mathbf{k}^{2} -
m^2(\mathbf{k},\Gamma)+\mathbf{q}^{2} +
\Pi_{A}(\mathbf{q},\Gamma)}{\sqrt{\mathbf{k}^{2} +
m^2(\mathbf{k},\Gamma)}}\left(\frac{\pi}{2} -
\arctan\left(\frac{\Gamma}{\sqrt{\mathbf{k}^{2} +
m^2(\mathbf{k},\Gamma)}}\right)\right)\right.\nonumber \\
&&\left.+\frac{\pi}{2}\frac{\Gamma^2+\mathbf{k}^{2} +
m^2(\mathbf{k},\Gamma)-\mathbf{q}^{2} -
\Pi_{A}(\mathbf{q},\Gamma)}{\sqrt{\mathbf{q}^{2} +
\Pi_{A}(\mathbf{q},\Gamma)}} -
\Gamma\ln\left(\frac{\Gamma^2+\mathbf{k}^{2} +
m^2(\mathbf{k},\Gamma)}{\mathbf{q}^{2} +
\Pi_{A}(\mathbf{q},\Gamma)}\right)\right]
\nonumber \\
&&\left.+\left[\Pi_{A}(\mathbf{q},\Gamma)\rightarrow
\Pi_{B}(\mathbf{q},\Gamma)\right]\right\}.
\label{Eq:GapEqFGammaNewAppr}
\end{eqnarray}
\begin{figure}[htbp]
\includegraphics[width=3in]{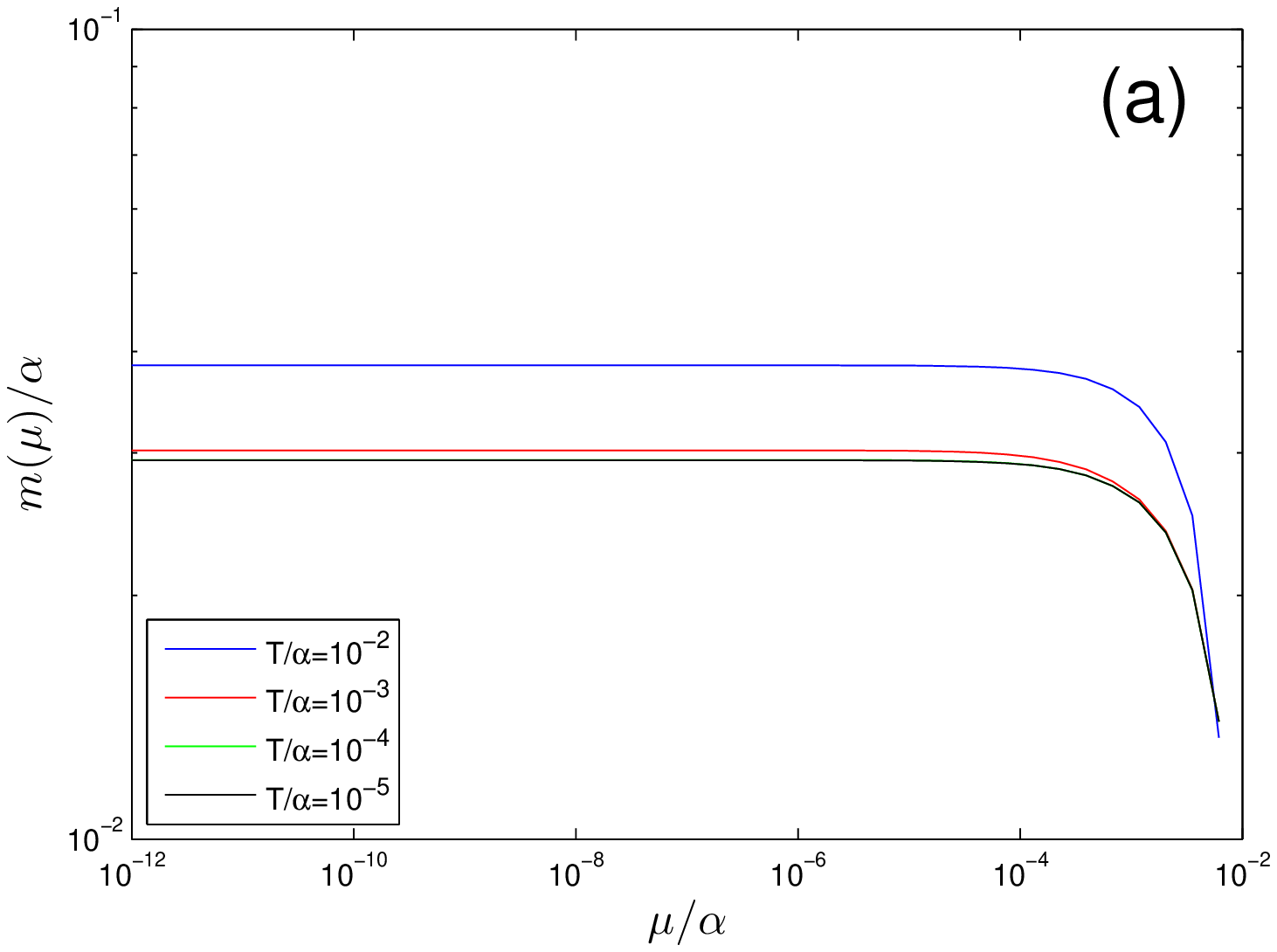}
\includegraphics[width=3in]{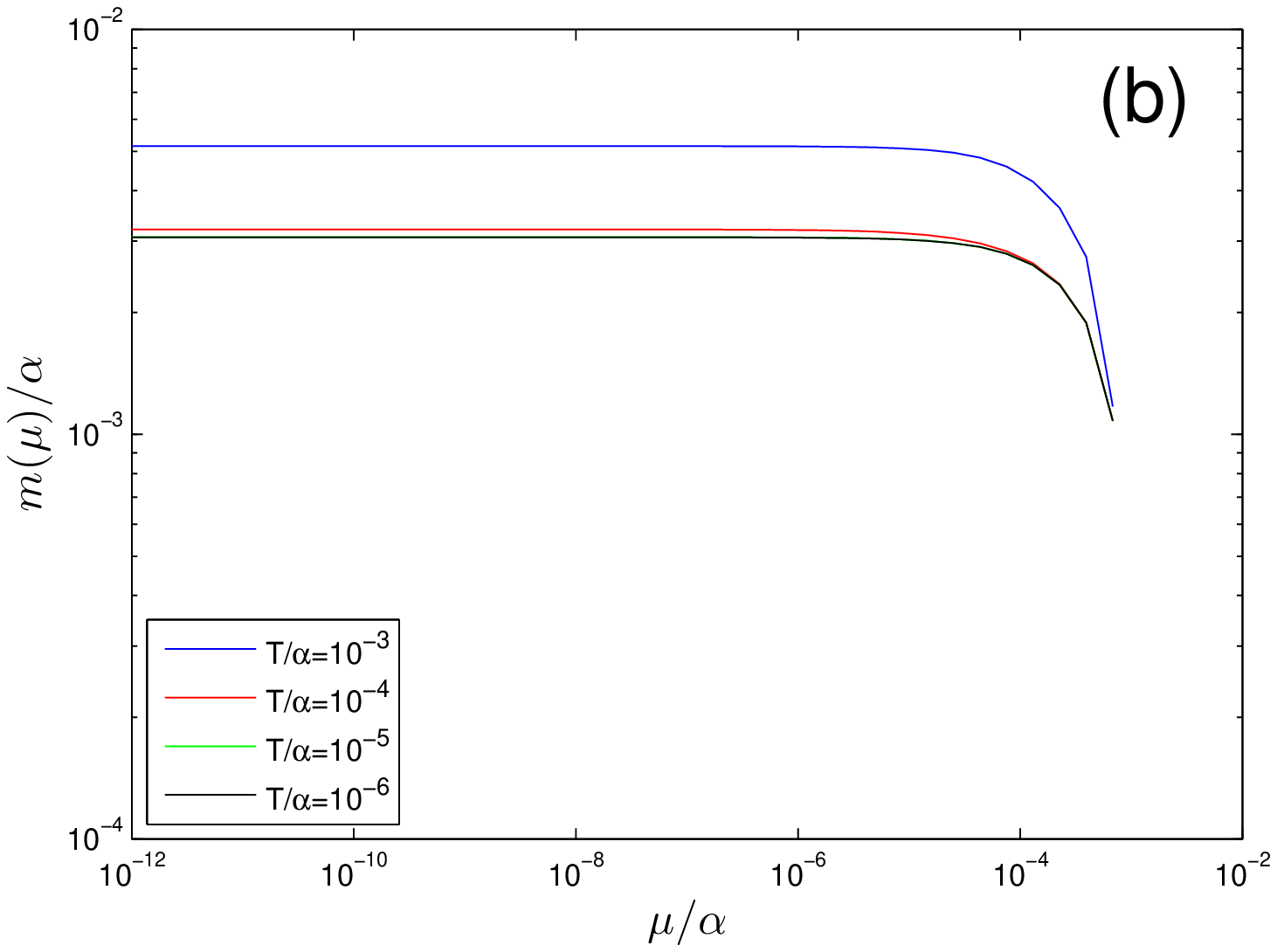}
\caption{Relation between $m(\mu)/\alpha$ and $\mu/\alpha$ for
different $T$ with $N=2,3$ in (a) and (b). Results are obtained
under the new approximation.}
\label{Fig:GapFTNewAppro}
\end{figure}
\begin{figure}[htbp]
\includegraphics[width=3in]{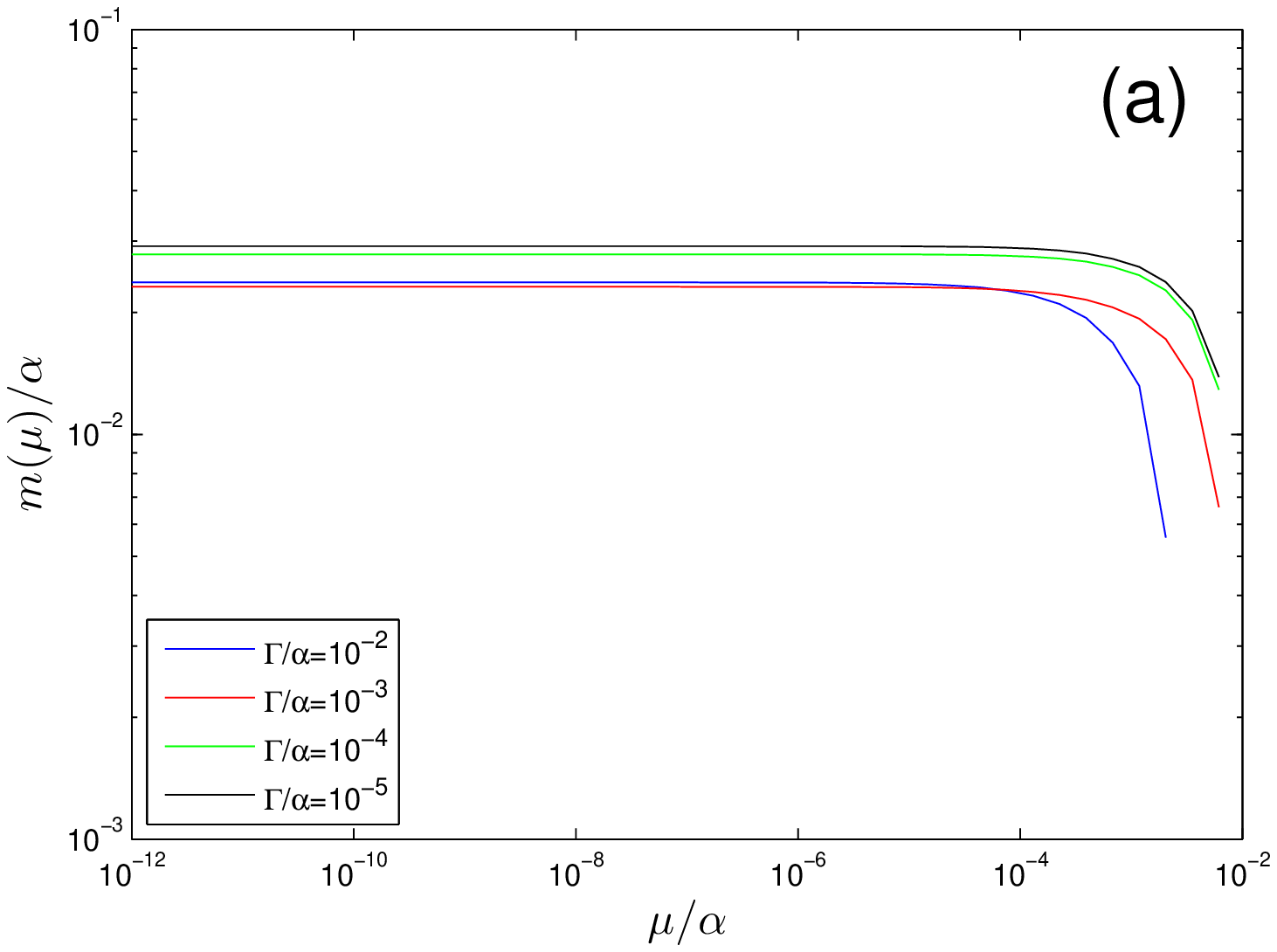}
\includegraphics[width=3in]{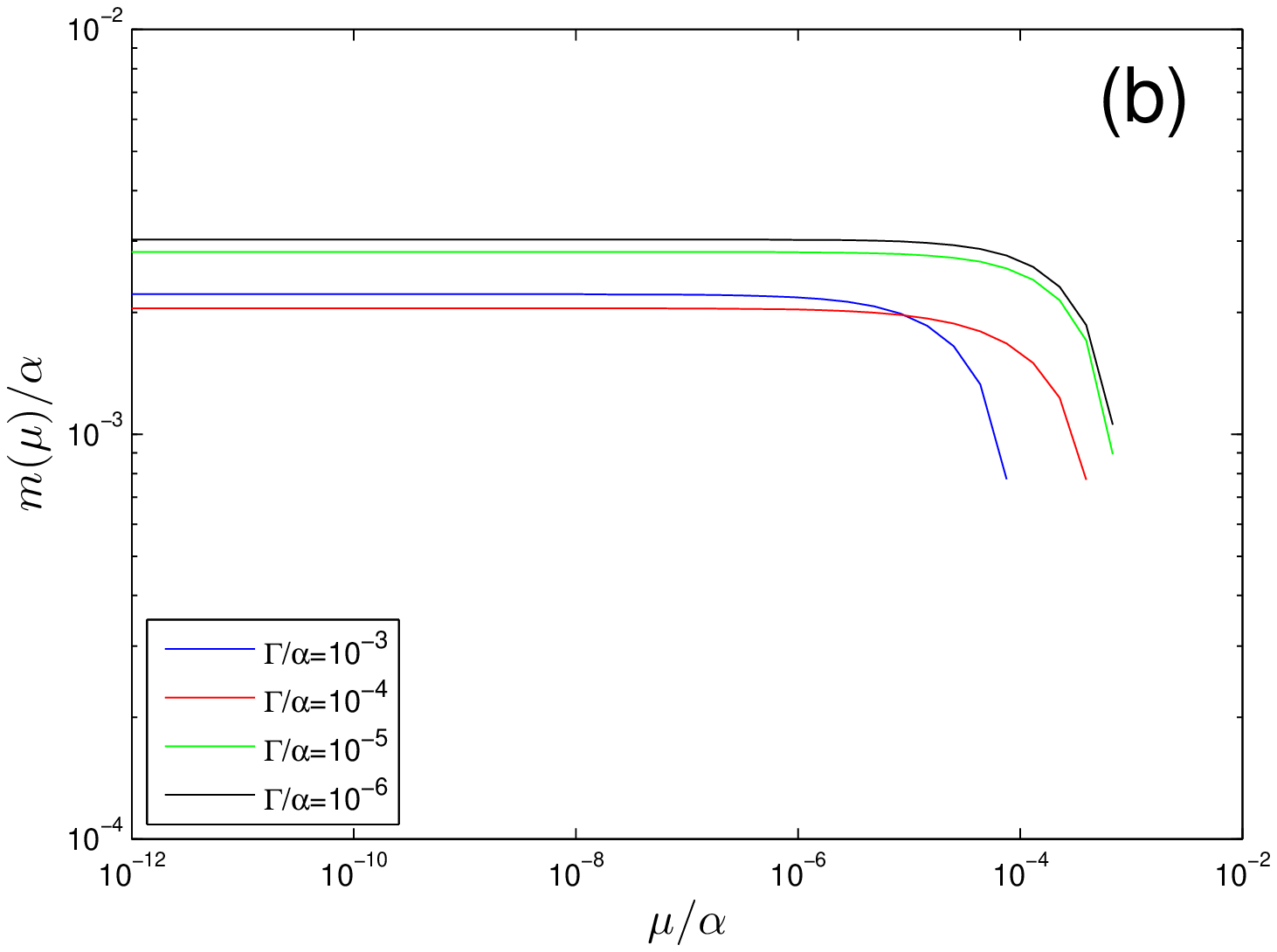}
\caption{Relation between $m(\mu)/\alpha$ and $\mu/\alpha$ for
different $\Gamma$ with $N=2,3$ in (a) and (b). Results are obtained
under the new approximation.}
\label{Fig:GapFGammaNewAppro}
\end{figure}
\end{widetext}

\begin{figure*}[htbp]
\center
\includegraphics[width=3in]{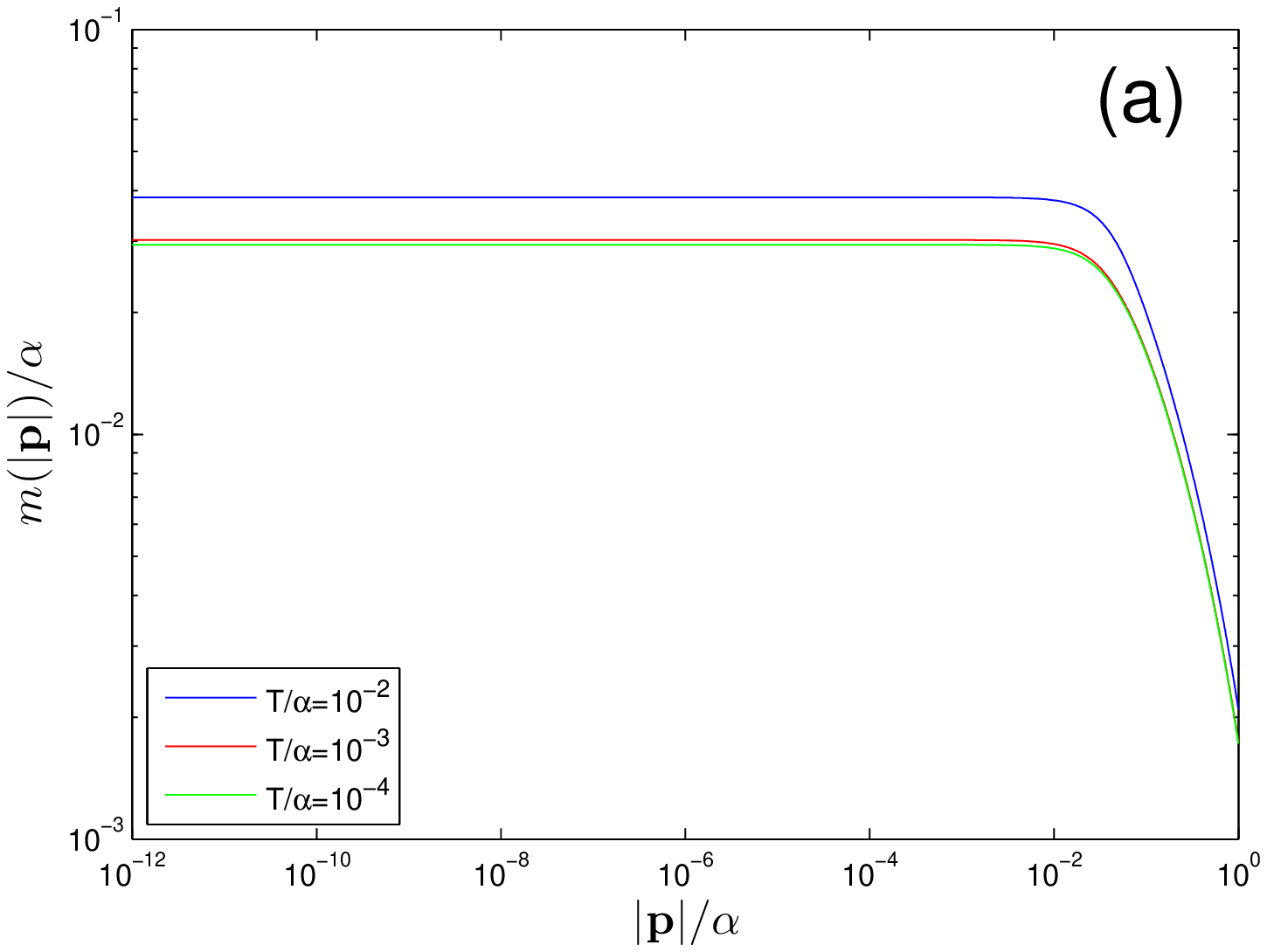}
\includegraphics[width=3in]{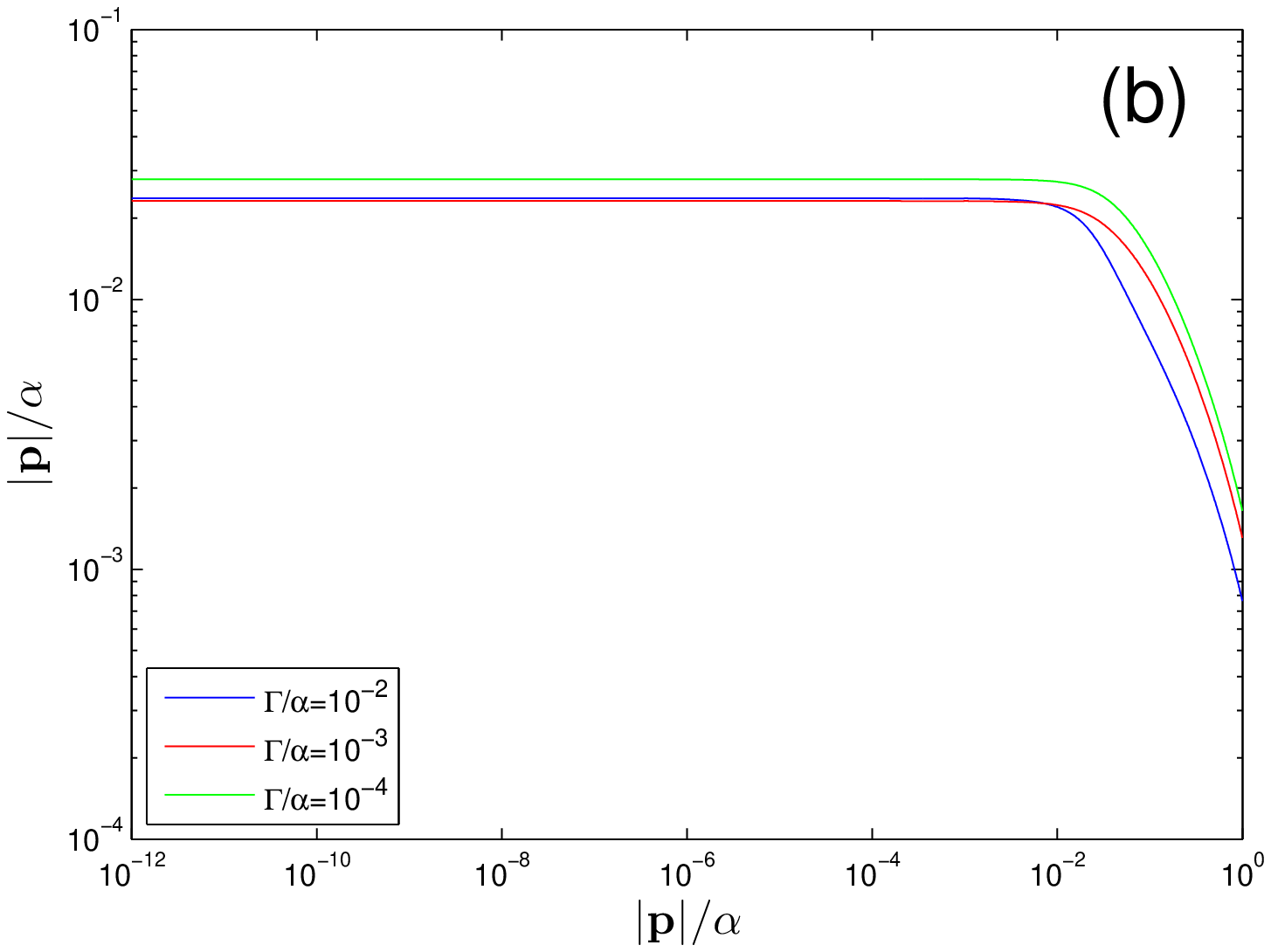}
\caption{Momentum dependence of $m(|\mathbf{p}|)$ under the new
approximation for various values of (a) $T$ and (b) $\Gamma$ with
$\mu=10^{-12}$.}\label{Fig:GapPFNewAppro}
\end{figure*}

Following the approach used in Sec.~\ref{Sec:GapInstan}, we
introduce an infrared cutoff $\mu$ to Eqs.~(\ref{Eq:GapEqFTNewAppr})
and (\ref{Eq:GapEqFGammaNewAppr}), and then solve these equations
numerically. In the case that $T \neq 0$ and $\Gamma = 0$, the
dependence of dynamical mass $m(\mu)$ on $\mu$ at finite $T$ with
$N=2$ and $N=3$ is shown in Figs.~\ref{Fig:GapFTNewAppro} (a) and
(b) respectively. As the infrared cutoff vanishes, $\mu \rightarrow
0$, $m(\mu)$ is saturated to certain finite values, which means the
infrared divergence encountered under the instantaneous
approximation does not exist under the new approximation. Analogous
calculations can be carried out in the case that $T = 0$ and $\Gamma
\neq 0$. We display the dependence of $m(\mu)$ on $\mu$ with $N=2$
and $N=3$ in Figs.~\ref{Fig:GapFGammaNewAppro}(a) and (b)
respectively. It is easy to see that $m(\mu)$ also approaches finite
values as $\mu \rightarrow 0$. Apparently, the new approximation
adopted in this section leads to convergent results for the
dynamical fermion mass, and is therefore more reliable than the
instantaneous approximation. We fix the infrared cutoff at
$\mu=10^{-12}$ and calculate the dynamical gap $m(|\mathbf{p}|)$
under the new approximation. The calculated $m(|\mathbf{p}|)$ under
the new approximation at different values of $T$ or different values of
$\Gamma$ is shown in Figs.~\ref{Fig:GapPFNewAppro} (a) and (b)
respectively.

Let us now analyze the origin of the infrared divergence. At $T \neq
0$ and $\Gamma = 0$, the DSE has the form
\begin{eqnarray}
m(p_{0},\mathbf{p},T) &=& \frac{e^{2}}{\beta}\sum_{q_{0} = 2n\pi
T}\int\frac{d^2\mathbf{q}}{(2\pi)^2}
\frac{m(k_{0},\mathbf{k},T)}{k_{0}^{2}+\mathbf{k}^{2} +
m^2(k_{0},\mathbf{k},T)}\nonumber \\
&&\times \left[ \frac{1}{q_{0}^2 +
\mathbf{q}^{2}+\Pi_{A}(q_{0},\mathbf{q},T)}\right.\nonumber \\
&&\left.+ \frac{1}{q_{0}^{2} + \mathbf{q}^2 +
\Pi_{B}(q_{0},\mathbf{q},T)}\right]. \label{Eq:GapTOriginal}
\end{eqnarray}
Since $m$ is finite after dynamical mass generation and
$\Pi_{A}(0,0,T)\propto T$ in the low energy limit, the terms
appearing in the first and second lines of the integral kernel are
safe in the infrared region. However, the term appearing in the
third line contains a potential infrared divergence. To make this
transparent, we divide the summation over $k_0$ as follows,
\begin{eqnarray}
I_{1}& \sim& \sum_{q_{0} = 2nT}\int\frac{d^2\mathbf{q}}{(2\pi)^{2}}
\frac{1}{q_{0}^{2}+\mathbf{q}^2+\Pi_{B}(q_{0},\mathbf{q},T)}
\nonumber \\
&\sim& \sum_{q_{0}=2nT(n\neq0)}\int\frac{d^2\mathbf{q}}{(2\pi)^{2}}
\frac{1}{q_{0}^{2}+\mathbf{q}^2+\Pi_{B}(q_{0},\mathbf{q},T)}
\nonumber \\
&&+\int\frac{d^2\mathbf{q}}{(2\pi)^{2}}\frac{1}{\mathbf{q}^2 +
\Pi_{B}(\mathbf{q},T)}.
\end{eqnarray}
Notice that $q_{0} = 2nT(n\neq0)$ is always finite at finite $T$
whenever $n\neq 0$, so the first term does not yield any infrared
divergence. On the contrary, the second term is dangerous because
$\Pi_{B}(\mathbf{q},T) = a_{1}\mathbf{q}^{2}$ with $a_{1} \propto
\frac{1}{T}$ for small momenta. We now simply focus on the potential
divergent term, and find that
\begin{eqnarray}
\int\frac{d^2\mathbf{q}}{(2\pi)^{2}}\frac{1}{\mathbf{q}^2 +
\Pi_{B}(\mathbf{q},T)} &\sim& \int\frac{d^2\mathbf{q}}{(2\pi)^{2}}
\frac{1}{c\mathbf{q}^2}\nonumber\\
&\sim& \frac{1}{c}\int_{\mu}^{\Lambda}
\frac{d|\mathbf{q}|}{|\mathbf{q}|}\sim \frac{1}{c}
\ln\left(\frac{\Lambda}{\mu}\right),\nonumber
\end{eqnarray}
where $c = 1+a_{1}$. It is clear that this term is divergent as the
infrared cutoff $\mu \rightarrow 0$. From the above analysis, we see
that the infrared divergence of DSE comes from the zero-energy
transfer processes mediated by the singular transverse component of
gauge boson propagator. Lee \cite{Lee98} noticed the existence of
infrared divergence, and then simply neglected the transverse
component of gauge boson propagator. The same strategy is widely
utilized in other works \cite{Dorey91, Dorey92, Aitchison92,
Aitchison94, Feng12A,Feng12B, Feng12C, Feng13, Yin14, Feng14}.

\begin{figure*}[htbp]
\center
\includegraphics[width=3in]{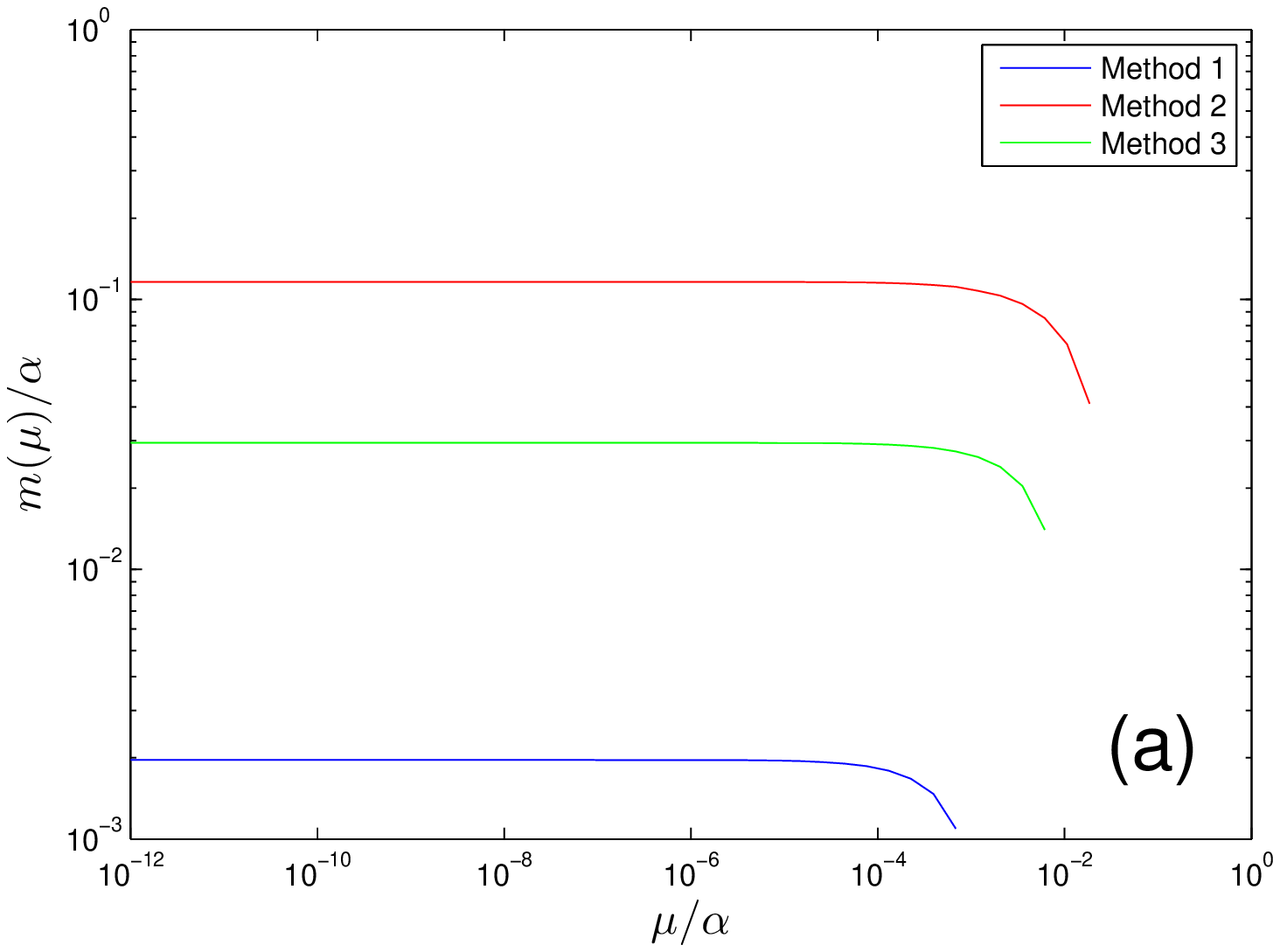}
\includegraphics[width=3in]{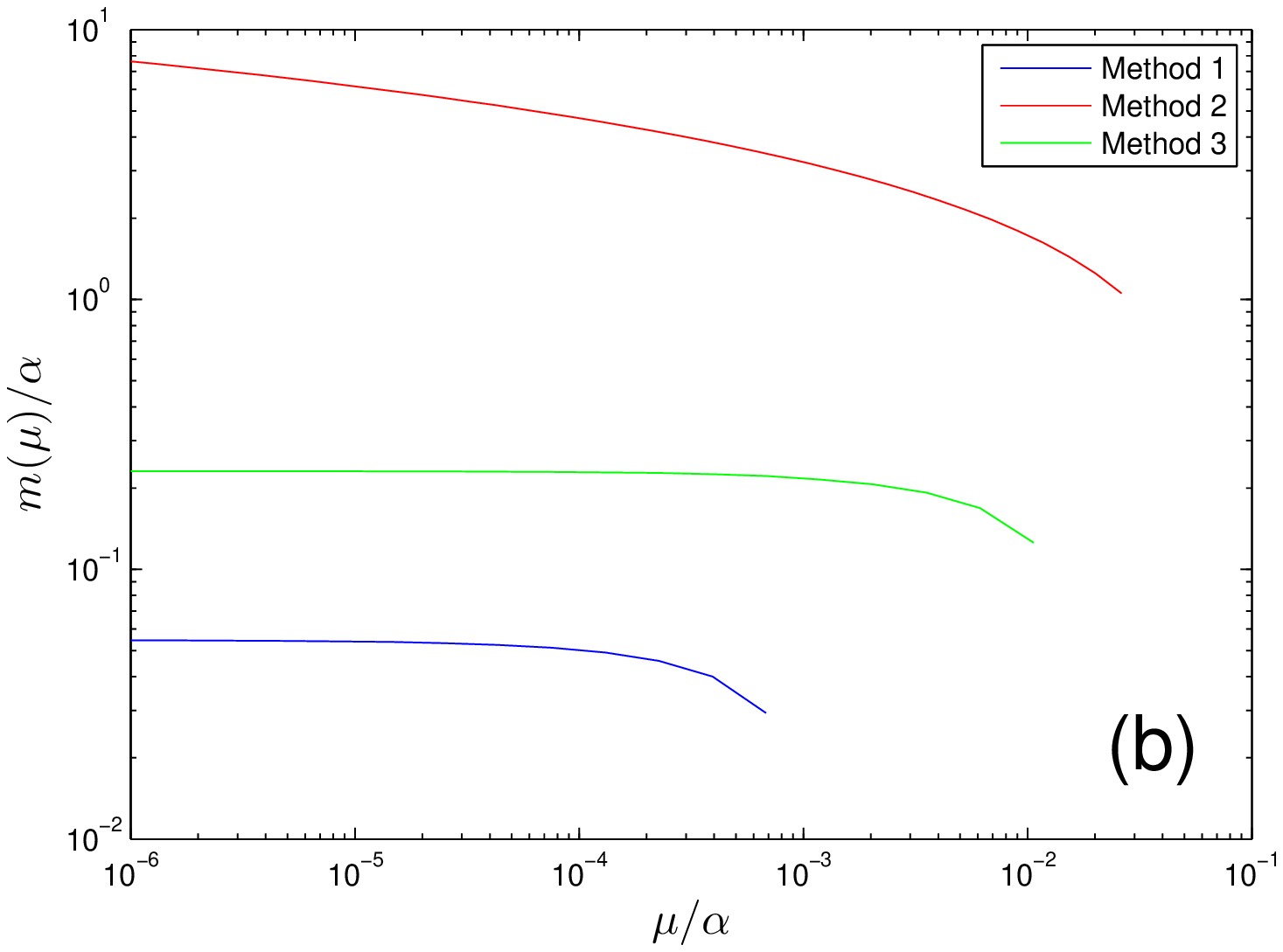}
\caption{Dependence of $m(\mu)/\alpha$ on $\mu/\alpha$ at $T = 0$
(a) neglecting and (b) including the feedback of $m(\mu)$ to the
polarization.} \label{Fig:GapZeroT}
\end{figure*}

More recently, Lo and Swanson \cite{Lo11} also stressed the
existence of infrared divergence and proposed to remove this
divergence by choosing an appropriate temperature dependent gauge
parameter. Their approach is basically equivalent to considering the
following DSE,
\begin{eqnarray}
m(p_{0},\mathbf{p},T) &=& \frac{e^{2}}{\beta}\int
\frac{d^2\mathbf{q}}{(2\pi)^2}
\frac{m(k_{0},\mathbf{k},T)}{k_{0}^{2}+\mathbf{k}^{2}
+ m^2(k_{0},\mathbf{k},T)}\nonumber \\
&& \times \left[\sum_{q_{0}=2n\pi T}
\frac{1}{q_{0}^2+\mathbf{q}^{2}+\Pi_{A}(q_{0},\mathbf{q},T)}\right.
\nonumber \\
&&\left.+\sum_{q_{0}=2n\pi T(n\neq0)}\frac{1}{q_{0}^{2}+\mathbf{q}^2
+ \Pi_{B}(q_{0},\mathbf{q},T)}\right],\nonumber
\end{eqnarray}
which ignores the zero frequency ($n = 0$) contribution of the
transverse component of gauge boson propagator. Careful numerical
computation of this equation is interesting, but challenging since
it is hard to sum over $n$ and at the same time integrate over
$\mathbf{q}$ with high precision. This is subjected to future
investigation.

Under the new approximation, the DSE of dynamical mass is simply
Eq.~(\ref{Eq:GapEqFTNewApprB}). The denominator of the kernel of
this equation contains a factor of $k_0^2 = \left((2n+1)\pi
T\right)^2$, whose minimum is $\pi^2 T^{2}$. We may consider this
term as an effective thermal mass of gauge boson, i.e.,
$m_\mathrm{a} \propto \pi^2T^2$, which serves as an infrared
regulator and eliminates the potential infrared divergence of the
dynamical fermion mass. This thermal mass exists only at finite $T$,
and vanishes naturally as $T \rightarrow 0$. To examine to what
extent the new approximation is valid, we will show in the next
section that Eq.~(\ref{Eq:GapEqFTNewApprB}) leads to results
qualitatively consistent with those obtained at $T = 0$
\cite{Appelquist88}.

We then turn to the case of finite fermion damping rate. At $T = 0$,
the DSE is given by Eq.~(\ref{Eq:GapEqFGammaNewApprB}). The
potential infrared divergence can only come from the second term in
the bracket. The most singular part is represented by
\begin{eqnarray}
I_{2}&\sim&\int\frac{dq_{0}}{2\pi}\int\frac{d^2\mathbf{q}}{(2\pi)^{2}}
\frac{1}{q_{0}^{2}+\mathbf{q}^2 +\Pi_{B}(q_{0},\mathbf{q},\Gamma)}.
\end{eqnarray}
In order to analyze potential infrared divergence, we replace
$\Pi_{B}(q_{0},\mathbf{q},\Gamma)$ with
$\Pi_{B}(\mathbf{q},\Gamma)$, then
\begin{eqnarray}
I_{2} &\sim& \int\frac{dq_{0}}{2\pi}\int
\frac{d^2\mathbf{q}}{(2\pi)^{2}}\frac{1}{q_{0}^{2} + \mathbf{q}^2 +
\Pi_{B}(\mathbf{q},\Gamma)}.
\end{eqnarray}
For small momenta, $\Pi_{B}(\mathbf{q},\Gamma)$ behaves as
$\Pi_{B}(\mathbf{q},\Gamma) = a_{2}\mathbf{q}^{2}$ with
$a_{2}\propto \frac{1}{\Gamma}$. It is clear that the potential
infrared divergence can be represented by
\begin{eqnarray}
I_{2} &\sim& \int\frac{dq_{0}}{2\pi}\int
\frac{d^2\mathbf{q}}{(2\pi)^{2}}\frac{1}{q_{0}^{2} + c'\mathbf{q}^2}
=\frac{1}{2\sqrt{c'}}\int\frac{d^2\mathbf{q}}{(2\pi)^{2}}
\frac{1}{|\mathbf{q}|}\nonumber \\
&=&\frac{1}{4\pi\sqrt{c'}}\int_{\mu}^{\Lambda} d|\mathbf{q}|
=\frac{1}{4\pi\sqrt{c'}}(\Lambda-\mu),
\end{eqnarray}
where $c'=1+a_{2}$. As $\mu \rightarrow 0$, $I_{2}$ is definitely
not divergent. This property should be fulfilled no matter what
approximation is used to calculate the DSE.

Under the instantaneous approximation, the DSE is represented by
Eq.~(\ref{Eq:GapFGammaInstanAppro}), which contains the following
singular contribution,
\begin{eqnarray}
I_{3} &\sim& \int\frac{d^2\mathbf{q}}{(2\pi)^{2}}
\frac{1}{\mathbf{q}^2 + \Pi_{B}(\mathbf{q},\Gamma)} = \frac{1}{2\pi
c'}\ln\left(\frac{\Lambda}{\mu}\right).
\end{eqnarray}
This function is divergent as $\mu \rightarrow 0$. Apparently, the
instantaneous approximation brings an artificial infrared divergence
that should not exist.

Under the new approximation proposed by us, the DSE is given by
Eq.(\ref{Eq:GapEqFGammaNewApprB}). The only possible singular part
can be simply written as
\begin{eqnarray}
I_{4} &\sim& \int\frac{dk_{0}}{2\pi}
\int\frac{d^2\mathbf{q}}{(2\pi)^{2}}\frac{1}{k_{0}^{2} +
\mathbf{q}^2 +\Pi_{B}(\mathbf{q},\Gamma)} \\
&\sim& = \frac{1}{4\pi\sqrt{c'}}(\Lambda-\mu),
\end{eqnarray}
which is not divergent and well consistent with Eq.~(51).

\section{Validity of new approximation\label{Sec:GapZero}}

In Sec.IV, we have adopted a new approximation to study the DSE for
dynamical fermion mass in the presence of finite temperature or
finite fermion damping. Compared to the popular instantaneous
approximation, the main advantage of the new approximation is that,
it retains both the longitudinal and transverse components of gauge
boson propagator, and at the same time leads to physically
meaningful, convergent results. To further see this point, it is now
interesting to make a more straightforward comparison between these
two approximations.

In this section, we consider zero temperature QED$_3$ in the clean
limit, namely $T = \Gamma = 0$. This model has already been
extensively investigated, and the dynamical mass obtained from DSE
is free of infrared divergence, which then can be considered as a
reference to examine the reliability of the results obtained at
finite temperature. For an approximation to be reliable, it should
work well at both zero and finite temperatures. As $T \rightarrow
0$, $m(T)$ oughts to approach a well-defined quantity $m(0)$, which
should be free of infrared divergence and as close in quantity as
possible to that obtained directly at $T = 0$. We now examine
whether the new approximation is valid according to this criterion.

If one neglects the feedback of fermion mass to the polarization
functions, the DSE to the lowest order of $1/N$-expansion is
known to be \cite{Appelquist88}
\begin{eqnarray}
m(p) = \frac{4\alpha}{N}\int\frac{d^3k}{(2\pi)^3}\frac{m(k)}{k^{2} +
m^2(k)}\frac{1}{q^{2} + \Pi(q)},\label{Eq:GapZeroTZeroGamma}
\end{eqnarray}
with $\Pi(q)=\alpha q$. This equation was first solved in
Ref.\cite{Appelquist88}, and the solution is very well known. Under
the instantaneous approximation, the DSE is simplified to
\begin{eqnarray}
m(\mathbf{p}) = \frac{8\alpha}{N}\int
\frac{d^2\mathbf{k}}{(2\pi)^2}
\frac{m(\mathbf{k})}{\sqrt{\mathbf{k}^{2} + m(\mathbf{k})}}
\frac{1}{\mathbf{q}^{2}+\Pi(\mathbf{q})},\label{Eq:GapZeroTZeroGammaInstan}
\end{eqnarray}
with $\Pi(|\mathbf{q}|)=\alpha |\mathbf{q}|$. Under the new
approximation, the corresponding DSE is
\begin{eqnarray}
m(\mathbf{p}) &=& \frac{4\alpha}{N}\int
\frac{d^2\mathbf{k}}{(2\pi)^2}
\frac{m(\mathbf{k})}{\sqrt{\mathbf{k}^{2}+m^2(\mathbf{k})}}
\frac{1}{\sqrt{\mathbf{q}^{2}+\Pi(\mathbf{q})}} \nonumber \\
&&\times\frac{1}{\sqrt{\mathbf{k}^{2} + m^2(\mathbf{k})} +
\sqrt{\mathbf{q}^{2} + \Pi(\mathbf{q})}},\label{Eq:GapZeroTZeroGammaNew}
\end{eqnarray}
where $\Pi(\mathbf{q})$ also equals to $\alpha |\mathbf{q}|$. The
dynamical fermion mass $m(\mu)$ as a function $\mu$ obtained in
three cases are shown in Fig.~\ref{Fig:GapZeroT} (a), represented by
lines with different colors. We notice that $m(\mu)$ is saturated to
finite values in all these three cases when $\mu \rightarrow 0$.
However, the fermion mass obtained in the new approximation is
closer to that obtained directly from
Eq.~(\ref{Eq:GapZeroTZeroGamma}) than the instantaneous
approximation. It seems that both of these two approximations lead
to convergent results for dynamical mass. Nevertheless, we still
need to examine whether these results are robust against higher
order corrections.

We then include the feedback of dynamical fermion mass to the
polarization. The polarization appearing in
Eq.~(\ref{Eq:GapZeroTZeroGamma}) becomes
\begin{eqnarray}
\Pi(q,m_{0}) &=& \frac{8\alpha q^2}{\pi}\left[\frac{m_0}{2q^{2}}
+ \frac{q^2 - 4m_0^2}{4q^{3} }\right.\nonumber \\
&&\left.\times\arcsin\left(\sqrt{\frac{q^2}{q^2 +
4m_0^2}}\right)\right],
\end{eqnarray}
and the polarization appearing in
Eqs.~(\ref{Eq:GapZeroTZeroGammaInstan}) and
(\ref{Eq:GapZeroTZeroGammaNew}) can be obtained by replacing $q$ of
$\Pi(q,m_0)$ with $|\mathbf{q}|$. In addition, $m_{0}$ is
substituted by $m(\mu)$. The dependence of dynamical mass $m(\mu)$
on $\mu$ is depicted in Fig.~\ref{Fig:GapZeroT}(b). We find that
$m(\mu)$ diverges as $\mu \rightarrow 0$ under the instantaneous
approximation. Nevertheless, $m(\mu)$ obtained in the other two
cases does not exhibit infrared divergence and remains finite as
$\mu \rightarrow 0$. These results further demonstrate that the
instantaneous approximation yields unphysical results, and that the
new approximation is more reliable.

\section{Summary and discussion\label{Sec:Conclusion}}

In this paper, we have studied dynamical fermion mass generation in
QED$_{3}$ after including the effects of finite temperature or
finite fermion damping rate. Many previous DSE calculations of
dynamical fermion mass adopted an instantaneous approximation, which
is often accompanied by simply ignoring the transverse component of
gauge interaction \cite{Dorey91, Dorey92, Aitchison92, Aitchison94,
Feng12A, Feng12B, Feng12C, Feng13, Yin14, Feng14}. As already
explained in the context, at finite temperature or at a finite
fermion damping rate, the longitudinal component of gauge
interaction becomes short-ranged due to static screening, whereas
the transverse component of gauge interaction remains long-ranged as
required by the local gauge invariance. It is therefore not
appropriate to ignore the more important contribution of gauge
interaction. However, we have showed that, if one adopts the
instantaneous approximation and meanwhile includes the complete
gauge boson propagator, the dynamical fermion mass exhibits infrared
divergence.

We have revisited this problem and employed a new approximation to
calculate the DSE for dynamical fermion mass. Under the new
approximation, both the longitudinal and transverse components of
gauge interaction are incorporated, and the results obtained from in
the DSE are free of infrared divergence. To further examine
the validity of the new approximation, we have also make a
comparison to the results obtained directly at zero temperature. In
summary, our calculations have showed that the new approximation
leads to more reliable results for the dynamical fermion mass than
the widely used instantaneous approximation.

The existence of infrared divergence is not special to the issue of
dynamical mass generation in finite temperature QED$_{3}$. Indeed,
similar divergence appears in a number of interacting gauge field
theories. For instance, Lee calculated the fermion damping rate
caused by gauge interaction within an effective non-relativistic
U(1) gauge field theory, and found non-Fermi liquid behavior at zero
temperature \cite{Lee92}. Nevertheless, the fermion damping rate
diverges at finite temperature \cite{Lee92}. Recently, analogous
divergence is also found in QED$_{3}$ defined at finite temperature
and finite chemical potential \cite{WangLiu10B}. We hope the
approach proposed and used in this paper could provide useful
insight into this problem.

Confinement is an important feature of QED$_{3}$. It is known from
previous studies \cite{Burden92, Maris95, Bashir08} that whether
this model is confining depends crucially on the behavior of the
polarization function in the low energy regime,
which is in turn determined by the dynamical fermion mass. It would
be interesting to apply the new approximation proposed here to
carefully calculate the polarization function at finite $T$ by including the impact
of dynamical fermion mass, and then to evaluate the critical
temperature for the confinement-deconfinement transition, following
the schemes presented in Refs.\cite{Burden92, Maris95, Bashir08}. It would also be
interesting to examine whether confinement and 
dynamical fermion mass generation take place simultaneously by analyzing the 
behavior of wave function renormalization 
\cite{Bashir08}. To address these issues, one needs to incorporate
the wave function renormalization and the vertex functions in the
DSEs, which are subjected to future research.

The authors acknowledge the financial support by the National
Natural Science Foundation of China under grants 11274286, 11174290,
and U1232142.

\appendix

\begin{widetext}
\section{Calculations of Polarization functions \label{Appendix:Polarization}}

In the Appendix, we present the detailed calculations for the
polarization functions in the presence of finite temperature $T$ and
finite fermion damping rate $\Gamma$. The calculations are performed
within the standard Matsubara formalism for finite temperature
quantum field theory.

\subsection{Expression for general $k_{0}$ and general $\Gamma$}

Starting from the effective fermion propagator given by Eq.~(\ref{Eq:GTGammaMass}), we
write the polarization functions $\Pi_{00}$ and $\Pi_{ii}$ in the
following forms
\begin{eqnarray}
\Pi_{00}(q_{0},\mathbf{q},T,m_0,\Gamma) &=& \frac{Ne^{2}}{\beta}
\sum_{n=-\infty}^{+\infty}\int\frac{d^{2}\mathbf{k}}{(2\pi)^{2}}
\mathrm{Tr}\left[G(k_{0},\mathbf{k}) \gamma_{0}G(k_{0} +
q_{0},\mathbf{k}+\mathbf{q})\gamma_{0}\right],\label{Eq:Pi00DefAppen1}
\\
\Pi_{ii}(q_{0},\mathbf{q},T,m_0,\Gamma) &=&\frac{Ne^{2}}{\beta}
\sum_{n=-\infty}^{+\infty}\int\frac{d^{2}\mathbf{k}}{(2\pi)^{2}}
\mathrm{Tr}\left[G(k_{0},\mathbf{k})\gamma_{i}G(k_{0} +
q_{0},\mathbf{k}+\mathbf{q})\gamma_{i}\right],\label{Eq:PiiiDefAppen1}
\end{eqnarray}
where $k_0 = (2n+1)\pi/\beta$ and $q_0 = 2\pi n'/\beta$ with $n$ and
$n'$ being integers, to the leading order of $1/N$-expansion.
Substituting Eq.~(\ref{Eq:GTGammaMass}) into
Eqs.~(\ref{Eq:Pi00DefAppen1}) and (\ref{Eq:PiiiDefAppen1}), and then
using the Feynman parametrization formula
\begin{equation}
\frac{1}{AB} = \int_{0}^{1}dx\frac{1}{[xA + (1-x)B]^2},
\end{equation}
we can get
\begin{eqnarray}
\Pi_{00}(q_{0},\mathbf{q},T,m_0,\Gamma) &=& \frac{4Ne^{2}}{\beta}
\int_{0}^{1}dx\int\frac{d^{2}\mathbf{l}}{(2\pi)^{2}}\left\{S_1-
2\left[\mathbf{l}^{2} + m_{0}^{2} + x(1-x)q_{0}^{2} + x\left(k_{0} +
\frac{3}{2}q_{0}\right)\delta\right]S_2\right.\nonumber \\
&&\left.+ \left[(1-2x)q_{0}+\delta\right]S^*\right\}, \\
\Pi_{ii}(q_{0},\mathbf{q},T,m_0,\Gamma) &=&-\frac{8Ne^{2}}{\beta}
\int_{0}^{1}dx\int\frac{d^{2}\mathbf{l}}{(2\pi)^{2}}\left\{S_1-
\left[\mathbf{l}^{2}+2x(1-x)\left(q_{0}^{2}+\frac{\mathbf{q}^{2}}{2}\right)
+ x\left(2k_{0}+3q_{0}\right)\delta\right]S_2\right. \nonumber \\
&&\left.+ \left[(1-2x)q_{0}+\delta\right]S^*\right\},
\end{eqnarray}
with
\begin{eqnarray}
S_{i} &=& \sum_{n=-\infty}^{\infty}\frac{1}{\left[l_{0}^{2} +
\mathbf{l}^{2}+m_{0}^{2}+x(1-x)q^2+2x\left(k_{0}+q_{0}\right)\delta\right]^{i}},
\\
S^{*} &=& \sum_{n=-\infty}^{\infty}\frac{l_{0}}{\left[l_{0}^{2} +
\mathbf{l}^{2}+m_{0}^{2}+x(1-x)q^2+2x\left(k_{0}+q_{0}\right)\delta\right]^{2}}.
\end{eqnarray}
where $l^{2}=l_{0}^{2}+\mathbf{l}^{2}$ with $l_{0} = k_{0}+xq_{0} +
\Gamma\mathrm{sgn}(k_{0})$, $q^2 = q_{0}^2 + \mathbf{q}^2$, and
$\delta = \Gamma\left[\mathrm{sgn}\left(k_{0} + q_{0}\right) -
\mathrm{sgn}(k_{0})\right]$. When $\delta \neq 0$, the frequency
summation cannot be carried out precisely. There are two ways to
make $\delta = 0$. First, $q_0 = 0$, corresponding to the static
limit. Second, $\Gamma = 0$, corresponding to the clean limit of the
system (without any disorder). Next we calculate the polarization
functions in these two cases respectively.

\subsection{Calculation of polarization functions in the limit $q_{0} = 0$}

For a general constant $\Gamma$, we have
\begin{eqnarray}
\Pi_{00}(\mathbf{q},T,m_0,\Gamma) &=&\frac{4Ne^{2}}{\beta}
\int_{0}^{1}dx\int\frac{d^{2}\mathbf{l}}{(2\pi)^{2}} \left[S_1 -
2\left(\mathbf{l}^{2}+m_{0}^{2}\right)S_2\right],\label{Eq:Pi00DefAppen2}
\\
\Pi_{ii}(\mathbf{q},T,m_0,\Gamma) &=&-\frac{8Ne^{2}}{\beta}
\int_{0}^{1}dx\int\frac{d^{2}\mathbf{l}}{(2\pi)^{2}}\left[S_1-
\left[\mathbf{l}^{2}+x(1-x)\mathbf{q}^2\right]S_2\right],\label{Eq:PiiiDefAppen2}
\end{eqnarray}
with
\begin{eqnarray}
S_{i}&=&\left(\frac{\beta}{2\pi}\right)^{2i} \sum_{n=0}^{\infty}
\frac{2}{\left[\left(n+\frac{1}{2}
+X\mathrm{sgn}\left(n+\frac{1}{2}\right)\right)^{2} +
Y^2\right]^{i}},
\end{eqnarray}
where $X=\frac{\beta}{2\pi}\Gamma$ and $Y =
\frac{\beta}{2\pi}\sqrt{\mathbf{l}^{2} + m_{0}^{2} +
x(1-x)\mathbf{q}^2}$. Summing over $n$, it is easy to get
\begin{eqnarray}
S_{1} = \frac{\beta^2}{2\pi^2Y} \mathrm{Im}\left[\psi
\left(\frac{1}{2}+X+iY\right)\right],
\end{eqnarray}
which then leads to
\begin{eqnarray}
S_{2} &=& -\frac{\beta^2}{8\pi^2Y}\frac{\partial S_{1}}{\partial Y}
=\frac{\beta^4}{16\pi^4Y^3} \mathrm{Im}\left[\psi\left(\frac{1}{2} +
X+iY\right)\right] - \frac{\beta^4}{16\pi^4Y^2}
\frac{\partial\mathrm{Im}
\left[\psi(\frac{1}{2}+X+iY)\right]}{\partial Y}.
\end{eqnarray}
Substituting the expressions of $S_{i}$ into
Eqs.~(\ref{Eq:Pi00DefAppen2}) and (\ref{Eq:PiiiDefAppen2}), we have
\begin{eqnarray}
\Pi_{00}(\mathbf{q},T,m_0,\Gamma)
&=&\frac{2Ne^{2}}{\pi^2}
\int_{0}^{1}dx\int_{\sqrt{m_{0}^{2}+C_{q}^{2}}}^{\Lambda}dt\left\{
\frac{C_{q}^2}{t^2} F_{1}(t,T,\Gamma)+\frac{t^2-C_{q}^{2}}{t}
\frac{\partial F_{1}(t,T,\Gamma)}{\partial t}\right\},
\\
\Pi_{ii}(\mathbf{q},T,m_0,\Gamma)
&=&-\frac{2Ne^{2}}{\pi^2}
\int_{0}^{1}dx\int_{\sqrt{m_0^2+C_q^2}}^{\Lambda}dt\left\{
\frac{t^{2}+m_{0}^{2}}{t^2}
F_{1}(t,T,\Gamma)+\frac{t^2-m_{0}^{2}}{t}
\frac{\partial\mathrm{Im}F_{1}(t,T,\Gamma)}{\partial t}\right\},
\end{eqnarray}
where $C_{q} = \sqrt{x\left(1-x\right)\mathbf{q}^2}$ and
$F_{1}(t,T,\Gamma) = \mathrm{Im}\left[\psi(\frac{1}{2} +
\frac{\Gamma}{2\pi T} + i\frac{t}{2\pi T})\right]$. Now we are
interested in the limiting behavior of $\Pi_{00}$ and $\Pi_{ii}$ at
zero temperature. As $T \rightarrow 0$, we know that
\begin{equation}
\lim_{T \rightarrow 0}
\mathrm{Im}\left[\psi\left(\frac{1}{2}+\frac{\Gamma+it}{2\pi
T}\right)\right] =\arctan\left(\frac{t}{\Gamma}\right).
\end{equation}
Therefore, at zero temperature the polarization functions can be
written as
\begin{eqnarray}
\Pi_{00}(\mathbf{q},m_0,\Gamma) &=& \frac{2Ne^{2}}{\pi^2}
\left\{\Gamma\ln\left(\frac{\Lambda}{\sqrt{\Gamma^{2}+m_{0}^{2}}}\right)
+ \Gamma \left[1+\frac{K_{2}}{2|\mathbf{q}|}
\ln\left(\frac{K_{2}-|\mathbf{q}|}{K_{2} +
|\mathbf{q}|}\right)\right] + \mathbf{q}^{2}\int_{0}^{1}
dx\frac{x(1-x)}{K_{1}}\arctan\left(\frac{K_{1}}{\Gamma}\right)\right\},
\nonumber \\
\Pi_{ii}(\mathbf{q},m_0,\Gamma) &=& \frac{2Ne^{2}}{\pi^2}
\mathbf{q}^2 \int_{0}^{1}dx\frac{x(1-x)}{K_{1}}\arctan
\left(\frac{K_{1}}{\Gamma}\right),
\end{eqnarray}
with $K_{1}=\sqrt{m_{0}^{2} + x(1-x)\mathbf{q}^2}$ and $K_{2} =
\sqrt{4\left(\Gamma^2+m_{0}^{2}\right) + \mathbf{q}^2}$.

\subsection{Calculation of polarization functions in the clean limit $\Gamma=0$}

For general $q_{0}$, we have
\begin{eqnarray}
\Pi_{00}(q_{0},\mathbf{q},T,m_0) &=&\frac{4Ne^{2}}{\beta}
\int_{0}^{1}dx\int\frac{d^{2}\mathbf{l}}{(2\pi)^{2}}\left\{S_1-
2\left[\mathbf{l}^{2} + m_{0}^{2} +
x(1-x)q_{0}^{2}\right]S_2+(1-2x)q_{0}S^*\right\},
\label{Eq:Pi00DefAppen3} \\
\Pi_{ii}(q_{0},\mathbf{q},T,m_0) &=& -\frac{8Ne^{2}}{\beta}
\int_{0}^{1}dx\int\frac{d^{2}\mathbf{l}}{(2\pi)^{2}}\left\{S_1-
\left[\mathbf{l}^{2} + 2x(1-x)\left(q_{0}^{2} +
\frac{\mathbf{q}^{2}}{2}\right)\right]S_2 +
(1-2x)q_{0}S^*\right\}.\label{Eq:PiiiDefAppen3}
\end{eqnarray}
with
\begin{eqnarray}
S_{i} &=& \left(\frac{\beta}{2\pi}\right)^{2i}
\sum_{n=-\infty}^{\infty}
\frac{1}{\left[\left(n+\frac{1}{2}+X\right)^{2} + Y^2\right]^{i}},
\\
S^{*} &=& \left(\frac{\beta}{2\pi}\right)^{3}
\sum_{n=-\infty}^{\infty}\frac{n+\frac{1}{2}+X}{\left[\left(n +
\frac{1}{2}+X\right)^{2} + Y^2\right]^{2}}.
\end{eqnarray}
where $X=\frac{\beta}{2\pi}xq_{0}$ and $Y =
\frac{\beta}{2\pi}\sqrt{\mathbf{l}^{2} +
m_{0}^{2}+x(1-x)\left(q_{0}^2+\mathbf{q}^2\right)}$. Carrying out
the frequency summation yields
\begin{eqnarray}
S_{1} &=& \frac{\beta^2}{4\pi^2Y}
\mathrm{Im}\left[\psi\left(\frac{1}{2}+X+iY\right)\right] +
\frac{\beta^2}{4\pi^2Y} \mathrm{Im}\left[\psi\left(\frac{1}{2} -
X+iY\right)\right].\label{Eq:S1ZeroGammaFiniteq0}
\end{eqnarray}
It is then straightforward to obtain
\begin{eqnarray}
S_{2} &=& -\frac{\beta^2}{8\pi^2Y}\frac{\partial S_{1}}{\partial Y}
=\frac{\beta^4}{32\pi^4Y^2}\left\{\frac{1}{Y}
\mathrm{Im}\left[\psi\left(\frac{1}{2}+X+iY\right)\right] -
\frac{\partial\mathrm{Im}\left[\psi(\frac{1}{2}+X+iY)\right]}{\partial
Y} + \frac{1}{Y} \mathrm{Im}\left[\psi\left(\frac{1}{2} -
X+iY\right)\right]\right.\nonumber \\
&&\left.-\frac{\partial\mathrm{Im}\left[\psi(\frac{1}{2} -
X+iY)\right]}{\partial Y}\right\},\label{Eq:S2ZeroGammaFiniteq0}
\\
S^{*} &=& -\frac{\beta}{4\pi}\frac{\partial S_{1}}{\partial X} =
-\frac{\beta^3}{16\pi^3Y}\left\{\frac{\partial
\mathrm{Im}\left[\psi(\frac{1}{2}+X+iY)\right]}{\partial X}
+\frac{\partial\mathrm{Im}\left[\psi(\frac{1}{2}-X+iY)\right]}{\partial
X}\right\}.\label{Eq:StarZeroGammaFiniteq0}
\end{eqnarray}
Substituting Eqs.~(\ref{Eq:S1ZeroGammaFiniteq0}),
(\ref{Eq:S2ZeroGammaFiniteq0}), and (\ref{Eq:StarZeroGammaFiniteq0})
into Eqs.~(\ref{Eq:Pi00DefAppen3}) and (\ref{Eq:PiiiDefAppen3}), the
polarization functions can be written as
\begin{eqnarray}
\Pi_{00}(q_{0},\mathbf{q},T,m_0) &=& \frac{Ne^{2}}{\pi^2}
\int_{0}^{1}dx\int_{\sqrt{m_{0}^{2}+C_{q}^{2}}}^{\Lambda}dt\left\{
\frac{B_{q}^2}{t^2}F_{2}(x,q_{0},t,T)+\frac{t^2-B_{q}^{2}}{t}
\frac{\partial\left(F_{2}(x,q_{0},t,T)\right)}{\partial
t}\right.\nonumber \\
&&\left.-\frac{\left(1-2x\right)q_{0}}{2} \frac{\partial
\left(F_{2}(x,q_{0},t,T)\right)}{\partial
(xq_{0})}\right\},\label{Eq:Pi00DefAppen4} \\
\Pi_{ii}(q_{0},\mathbf{q},T,m_0) &=& -\frac{Ne^{2}}{\pi^2}
\int_{0}^{1}dx\int_{\sqrt{m_{0}^{2}+C_{q}^{2}}}^{\Lambda}dt t
\left\{\frac{t^{2} - B_{q}'^2 +
m_{0}^{2}}{t^2}\left(F_{2}(x,q_{0},t,T)\right) +
\frac{t^2+B_{q}'^{2}-m_{0}^{2}}{t}
\frac{\partial\left(F_{2}(x,q_{0},t,T)\right)}{\partial
t}\right.\nonumber \\
&&\left.-\left(1-2x\right)q_{0}\frac{\partial
\left(F_{2}(x,q_{0},t,T)\right)}{\partial
(xq_{0})}\right\},\label{Eq:PiiiDefAppen4}
\end{eqnarray}
where $B_{q} = \sqrt{x(1-x)\mathbf{q}^2}$, $B_{q}' =
\sqrt{x(1-x)q_{0}^2}$, and $t = \sqrt{\mathbf{l}^{2} +
m_{0}^{2}+x(1-x)\left(q_{0}^{2} + \mathbf{q}^2\right)}$. Here,
$F_{2}(x,q_{0},t,T) =
\mathrm{Im}\left[\psi(\frac{1}{2}+\frac{xq_{0}}{2\pi T} +
i\frac{t}{2\pi T})\right] +
\mathrm{Im}\left[\psi(\frac{1}{2}-\frac{xq_{0}}{2\pi T} +
i\frac{t}{2\pi T})\right]$. Since $\psi(1-z) = \psi(z) + \pi\cot(\pi
z)$, we have
\begin{eqnarray}
F_{2}(x,q_{0},t,T) &=& \frac{\pi}{2i}
\left[-\tan\left(\frac{xq_{0}}{2 T} - i\frac{t}{2 T}\right) +
\tan\left(\frac{xq_{0}}{2 T} + i\frac{t}{2
T}\right)\right].\label{Eq:F2Appen}
\end{eqnarray}
Substituting Eq.~(\ref{Eq:F2Appen}) into
Eqs.~(\ref{Eq:Pi00DefAppen4}) and (\ref{Eq:PiiiDefAppen4}), the
polarization functions become
\begin{eqnarray}
\Pi_{00}(q_{0},\mathbf{q},T,m_0) &=& \frac{Ne^{2}}{2\pi }
\int_{0}^{1}dx\left\{\frac{2}{\beta}\ln\left(4\left[\cosh^2
\left(\frac{1}{2}\beta K_{3}\right) - \sin^2\left(\frac{1}{2}\beta
xq_{0}\right)\right]\right)
\right.\nonumber \\
&&\left.-\frac{1}{K_{3}}\frac{\left[m_0^2 +
x(1-x)q_{0}^{2}\right]\sinh(\beta K_{3})}
{\cosh^2(\frac{1}{2}\beta K_{3}) - \sin^2(\frac{1}{2}\beta
xq_{0})} - \frac{1}{2}\frac{(1-2x)q_{0}\sin(\beta
xq_{0})}{\cosh^2(\frac{1}{2} \beta K_{3}) -
\sin^2(\frac{1}{2}\beta xq_{0})}\right\},
\\
\Pi_{ii}(q_{0},\mathbf{q},T,m_0) &=&\frac{Ne^{2}}{2\pi }
\int_{0}^{1}dx\left\{\frac{1}{K_{3}}\frac{x(1-x)\left(2q_{0}^{2} +
\mathbf{q}^{2}\right)\sinh(\beta K_{3})} {\cosh^2(\frac{1}{2}
\beta K_{3}) - \sin^2(\frac{1}{2}\beta xq_{0})} +
\frac{(1-2x)q_{0}\sin(\beta xq_{0})}{\cosh^2(\frac{1}{2}\beta
K_{3}) -\sin^2(\frac{1}{2}\beta xq_{0})}\right\},
\end{eqnarray}
with $K_{3} = \sqrt{m_0^2 + x(1-x) \left(q_{0}^{2} +
\mathbf{q}^2\right)}$.

At zero temperature, the polarization functions are simplified to
\begin{eqnarray}
\Pi_{00}(q_{0},\mathbf{q},m_{0}) &=& \frac{Ne^{2}\mathbf{q}^2}{\pi}
\left[\frac{m_0}{2q^{2}} + \frac{q^2 - 4m_0^2}{4q^{3}}
\arcsin\left(\frac{q}{\sqrt{q^2 + 4m_0^2}}\right)\right],
\\
\Pi_{ii}(q_{0},\mathbf{q},m_{0}) &=& \frac{Ne^{2}\left(2q_{0}^{2} +
\mathbf{q}^2\right)}{\pi} \left[\frac{m_0}{2q^{2}} + \frac{q^2 -
4m_0^2}{4q^{3}}\arcsin\left(\frac{q}{\sqrt{q^2 +
4m_0^2}}\right)\right].
\end{eqnarray}
In the limit $q_{0}=0$, the polarization functions are
\begin{eqnarray}
\Pi_{00}(\mathbf{q},T,m_0) &=& \frac{Ne^{2}}{\pi }
\int_{0}^{1}dx\left\{2T
\ln\left[2\cosh\left(\frac{K_{1}}{2T}\right)\right] -
\frac{m_0^2}{K_{1}}\tanh\left(\frac{K_{1}}{2T}\right)\right\},
\nonumber \\
\Pi_{ii}(\mathbf{q},T,m_0) &=& \frac{Ne^{2}}{\pi }
\int_{0}^{1}dx\frac{x(1-x)\mathbf{q}^{2}}{K_{1}}
\tanh\left(\frac{K_{1}}{2T}\right).
\end{eqnarray}
\end{widetext}

\end{document}